\PassOptionsToPackage{dvipsnames,table,xcdraw}{xcolor}
\documentclass[10pt,journal,compsoc]{secs/IEEEtran}

\ifCLASSOPTIONcompsoc
  \usepackage[nocompress]{cite}
\else
  \usepackage{cite}
\fi

\usepackage{booktabs} %
\usepackage{graphicx}
\usepackage{epstopdf}
\usepackage{multirow}
\usepackage{subfigure}
\usepackage{array}
\usepackage{multicol}
\usepackage{xcolor}
\usepackage{amsfonts}
\usepackage{amsmath}
\renewcommand\fbox{\fcolorbox{orange}{white}}
\usepackage[export]{adjustbox}
\usepackage{tikz}
\usetikzlibrary{calc,fit,arrows.meta}
\newlength\aelength

\AppendGraphicsExtensions{.gif}

\usepackage{utils/my}

\graphicspath{{imgs}}
\usepackage[dvipsnames]{xcolor}
\definecolor{yjcolor}{rgb}{0.5,0.67,0.89}
\definecolor{zbtcolor}{rgb}{0.5,0.0,0.89}
\newcommand{\yl}[1]{{\color{black} #1}}

\newcommand{\zbt}[1]{{\color{black} #1}}
\newcommand{\hb}[1]{{\color{black}            {#1}}}

\usepackage[numbers]{natbib}

\usepackage{caption}

\ifCLASSINFOpdf

\begin{document}
\title{Mesh-based Gaussian Splatting for Real-time Large-scale Deformation}

\author{
       Lin~Gao,
       Jie~Yang,
       Bo-Tao~Zhang,
        Jia-Mu~Sun, 
        Yu-Jie~Yuan,
        Hongbo~Fu
        and Yu-Kun~Lai%
\IEEEcompsocitemizethanks{
\IEEEcompsocthanksitem Lin Gao,  Bo-Tao Zhang, Jia-Mu Sun and Yu-Jie Yuan are with the Beijing Key Laboratory of Mobile Computing and Pervasive Device, and also with the University of Chinese Academy of Sciences, Beijing, China.
\IEEEcompsocthanksitem Jie Yang is with the Beijing Key Laboratory of Mobile Computing and Pervasive Device and  Chinese Academy of Sciences, Beijing, China.
\protect\\
E-mail: \{gaolin, yangjie01, zhangbotao, sunjiamu21s, yuanyujie\}@ict.ac.cn
\IEEEcompsocthanksitem Hongbo Fu is with City University of Hong Kong.%
\protect\\
E-mail: fuplus@gmail.com
\IEEEcompsocthanksitem Yu-Kun Lai is with the School of Computer Science \& Informatics, Cardiff University, U.K.
\protect\\
E-mail: LaiY4@cardiff.ac.uk
}%
\thanks{Manuscript received April 19, 2023; revised August 26, 2023.}}

\markboth{IEEE Transactions on Pattern Analysis and Machine Intelligence,~Vol.~xx, No.~xx, June~2021}%
{Mesh-based Gaussian Splatting for Real-time Large-scale Deformation}

\IEEEtitleabstractindextext{%
\footnotesize
Neural implicit representations, including Neural Distance Fields and Neural Radiance Fields, have demonstrated significant %
\yl{capabilities for}
reconstructing %
\yl{surfaces with complicated \hb{geometry and} topology},
and %
\yl{generating}
novel views of a scene. 
Nevertheless, it is challenging for users to directly deform or manipulate these implicit representations with large deformations in the real-time fashion.%
Gaussian Splatting \yl{(GS)}
has recently become a promising method with explicit geometry for representing static scenes and facilitating high-quality and real-time synthesis of novel views. 
\yl{However, it cannot be easily deformed due to the use of discrete Gaussians and lack of explicit topology}.
\yl{To address this,}
we %
\yl{develop a novel GS-based method that enables interactive deformation.
}
Our %
\yl{key} idea is to design an innovative mesh-based \hb{GS} %
representation, which is integrated into \yl{Gaussian} learning and \hb{manipulation}. %
\hb{3D Gaussians are defined over an explicit mesh, and they} 
\yl{are}
bound with each other: the rendering of 3D Gaussians guides the mesh face split \yl{for adaptive refinement}, 
and the mesh face split %
\yl{directs}
the \yl{splitting} of 3D \yl{Gaussians}. 
\yl{Moreover, the explicit mesh constraints help regularize \hb{the} Gaussian distribution, %
\yl{suppressing poor\hb{-}quality}
Gaussians (\eg misaligned Gaussians, long-narrow shaped Gaussians), thus enhancing %
visual quality and avoiding artifacts during deformation.}
\yl{Based on this representation, we further introduce a large-scale Gaussian deformation technique} to enable %
deformable \hb{GS}, %
which \yl{alters} the parameters of 3D \yl{Gaussians} according to the \yl{manipulation of the associated mesh.} \yl{Our method
benefits from existing mesh deformation datasets for more realistic data-driven Gaussian deformation.}
\yl{Extensive} experiments show that our approach \yl{achieves high-quality reconstruction and effective deformation,}
while maintaining the promising rendering results at a high \hb{frame rate} %
(65 FPS \yl{on average} \yl{on a single commodity GPU}).

}

\maketitle

\IEEEdisplaynontitleabstractindextext

\IEEEpeerreviewmaketitle

\begin{figure*}
\centering
{
\begin{overpic}[width=\linewidth,trim={0cm 0.45cm 0cm 0cm},clip]{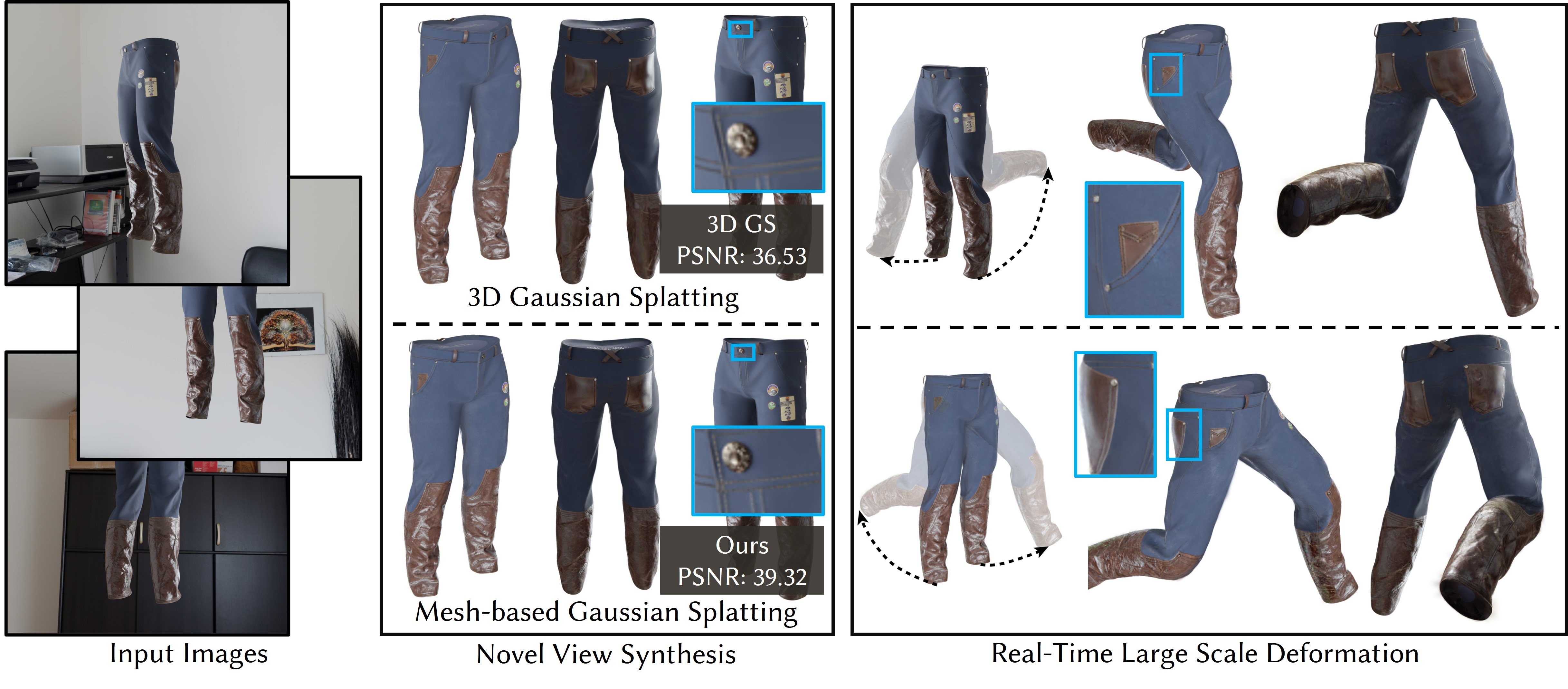}
\end{overpic}
}\caption{\yl{Given a set of \hb{multi-view images of an object} %
(left), we reconstructs the object \hb{with} %
the proposed Mesh-based Gaussian Splatting representation, involving both 3D Gaussians and an associated mesh. %
the mesh is adaptive refined along with Gaussian \yl{splitting},
and also served as effective regularization. As a result, our method achieves %
\hb{higher}-quality novel view synthesis, compared with 3D Gaussian splatting~\cite{kerbl20233d}. 
\yl{our} 3D Gaussian deformation method produces high quality deformation results in the real-time manner with large scale deformations. }
}
\label{fig:teaser}
\end{figure*}

\setlength{\abovedisplayskip}{3pt}
\setlength{\belowdisplayskip}{3pt}

\section{Introduction}\label{sec:intro}

Shape representations are fundamental in computer graphics and geometry processing. 
Traditionally, \yl{explicit} representations like point \yl{clouds}~\cite{qi2017pointnet,qi2017pointnet++,fan2017point,achlioptas2018learning}, voxels~\cite{maturana2015voxnet,choy20163d,3dgan}, \yl{meshes}~\cite{wang2018pixel2mesh,groueix2018atlasnet} are commonly used in various contexts due to their intuitive and deformation friendly property.
In recent years, 
Implicit \yl{representations} (\eg \yl{Neural Radiance Fields (NeRFs)}~\cite{mildenhall2021nerf,barron2022mip}, \yl{Signed Distance Fields (SDFs)~\cite{mescheder2019occupancy,chibane2020implicit,chen2019learning}}) \yl{have} drawn much attention since they are able to reconstruct highly realistic appearance and complicated geometry from only a few multi-view images. However, implicit representations also bear inherent drawbacks like slow rendering speed, limiting their applicability to %
\yl{practical}
applications. To overcome these drawbacks while preserving the advantages, 3D Gaussian Splatting (3DGS)~\cite{kerbl20233d} is proposed. 3DGS learns the spatial distributions of \yl{Gaussian} kernels from the initialization of SFM (Structure from Motion)~\cite{ullman1979interpretation} \yl{points}, which naturally provide \yl{an} explicit discrete 3D scene representation in contrast to the continuous representation used by NeRF. 3DGS has a small training cost and \zbt{can achieve high-quality real-time rendering based on differentiable rasterization}. 

Since 3DGS is built upon discrete Gaussian kernels, it \yl{may seem} natural to directly perform deformation on them. However, simple deformation methods may produce suboptimal results. For example, SC-GS~\cite{huang2023sc} learns sparse control points to \yl{model} the dynamics of 3D scenes, \hb{thus enabling} motion \yl{deformation} by manipulating the sparse points. Nevertheless, methods based on sparse control points \yl{struggle with complicated geometry or deformation}, due to the lack of topology prior. On the other hand, SuGaR~\cite{guedon2023sugar} employs a slightly more sophisticated pipeline. It \yl{extracts} \yl{an} explicit mesh from \yl{a} 3DGS representation by regularizing the \yl{Gaussians} to distribute over the surface, and can edit the 3DGS by manipulating the mesh.
However, \yl{when performing} deformation, SuGaR only adjusts the scale and rotation of existing Gaussian kernels without further merging or splitting \yl{them}, and mesh properties like \yl{normals} are also not taken into account. This can result in \yl{artifacts especially} when performing large-scale deformation, 
\yl{which} 
can greatly change the shape of the mesh. Simply deforming the Gaussians without any topology information generates strong misalignment artifact when performing large-scale deformation, as shown in the `Baseline' method in Fig.~\ref{fig:comparisons}. 

Motivated by the above observations, 
Our purposed method enables \yl{high-quality} real-time large-scale deformation on 3D Gaussian Splatting, as illustrated in Figure \ref{fig:teaser}. 
\yl{Our core} idea is to design an innovative mesh-based \yl{Gaussian} splatting representation, which is integrated into \yl{Gaussian} distribution learning and manipulations.  
\yl{In particular,} given a mesh of the scene extracted by an off-the-shelf method~\cite{wang2023neus2}, we bind the mesh and a 3DGS representation with each other. We leverage this binding to provide guidance for both training of 3DGS and deformation in a novel manner.
During the 3DGS learning process, its \yl{Gaussian} splitting is limited to two options: the first is splitting along the face, which means the two split kernels remain on the surface, \yl{and}
\hb{the other} %
is splitting along the normal, which means the two split kernels are aligned along the surface normal.
We observe this approach is more conducive to form manipulations and \yl{enhance} the rendering effects (\eg high-frequency details) by eliminating irrational \yl{Gaussians} (\eg misaligned \yl{Gaussians}, long-narrow shaped \yl{Gaussians}), \yl{which can lead to artifacts}.

Based on our proposed mesh-based GS,
we \yl{introduce a} large-scale \yl{Gaussian} deformation technique to \yl{achieve} editable GS, which \yl{alters} the parameters of 3D \yl{Gaussians} according to the mesh deformation.
In particular, we employ existing mesh deformation techniques \cite{gao2019sparse} on the mesh and apply the deformation gradients to the parameters of \yl{neighboring} Gaussians. This process can be directly rendered by the splatting procedure in \yl{real time}. This can equip our deformation with intuitive data-driven deformation which was only available in mesh-based methods before, thanks to the mesh-GS binding.
Additionally, a regularization loss is introduced to enforce the spatial continuity and local rationality of Gaussian \yl{shapes} and avoid the blurry visual artifacts due to the anisotropy of 3D \yl{Gaussian kernels} for \yl{Gaussian} deformation. 
Finally, to enable interactive deformation of 3D Gaussians, we design an interactive tool, which allows for real-time Gaussian manipulation and high-quality splatting, while adhering to user-friendly constraints.

Extensive experiments and ablations on public datasets and self-captured scenes demonstrate that \yl{our mesh-based GS achieves better novel view synthesis compared with existing techniques, while maintaining promising rendering speed (65 FPS on average on a single NVIDIA RTX 4090 GPU), and our method enables large-scale deformation %
of Gaussian splatting, outperforming existing methods.}
More results for real-time deformation are included in our video. 
Our contributions can be summarized as follows:
\begin{itemize}
	\item We propose a novel mesh-based Gaussian Splatting representation, which binds 3D Gaussians with the mesh representation and fully utilizes the mesh to guide the splitting of 3DGS, improving the quality of learned GS.
	\item With the proposed Gaussian representation, we introduce a large-scale Gaussian deformation method\hb{, which makes use of not only the vertex positions, but also deformation gradients to guide the GS.} It %
 takes the advantage of mesh deformation methods while preserving real-time rendering and high quality appearance robustly even when deformed at large scale.
	\item 
 Extensive experiments demonstrate that our method achieves superior performance on the efficiency and quality of deformation compared to existing methods.
\end{itemize}

\section{Related Work}\label{sec:related}

\subsection{3D Shape Representations}
\noindent\textbf{Explicit Representations}
\yl{Explicit representations dominated} the industries and academic research \yl{for} a long time. %
\hb{Classic representations, including} 
point clouds~\cite{qi2017pointnet,fan2017point}, voxels~\cite{maturana2015voxnet,choy20163d,3dgan,wang2018adaptive}, \yl{meshes}~\cite{hanocka2019meshcnn,groueix2018atlasnet,wang2018pixel2mesh,gaosdmnet2019,yang2022dsg}, 
\hb{have been revisited for 3D deep learning.}
Although 3D explicit \yl{representations have a} clear description on the geometry and appearance,
\yl{they %
\hb{lack a} %
flexible underlying topology representation and \hb{have} limited capabilities of representing realistic appearance.}

\noindent\textbf{Implicit \yl{Representations}}
Different from explicit \yl{representations}, 
\hb{implicit representations}, 
including \yl{signed} distance fields \yl{(SDFs)}~\cite{mescheder2019occupancy,chibane2020implicit,chen2019learning} \yl{and} unsigned distance fields \yl{(UDFs)}~\cite{chibane2020neural,guillard2021meshudf,liu2023neudf} \hb{can accurately model arbitrary geometry and topology.} \yl{Thanks} to the continuous nature of implicit representations, \yl{they} can be combined with neural \yl{networks} to support data-driven geometry learning.

In recent years, \yl{Neural Radiance Field} (NeRF) \yl{has} become increasingly popular as it allows for 3D optimization with only 2D supervision via volumetric rendering~\cite{kajiya1984ray}.
\hb{It}
has become prevalent in numerous tasks, such as 3D reconstruction~\cite{li2023neuralangelo,wang2021neus}, 3D generation~\cite{poole2022dreamfusion,liu2023zero}, and editing~\cite{haque2023instruct,wang2023seal,liu2021editing}. %
Nevertheless, \yl{implicit} approaches suffer from extensive sampling to fit \hb{implicit functions of 3D scenes}. %
This leads to significant computational costs, particularly in high-resolution or interactive rendering scenarios, %
\yl{even with}
accelerated NeRF versions~\cite{muller2022instant,yu2021plenoctrees,fridovich2022plenoxels}. \yl{It} is difficult to achieve real-time rendering and high quality view synthesis at the same time.
\noindent\textbf{Gaussian Splatting}
Recently, 3DGS has emerged as a \yl{viable} alternative 3D representation to NeRF, demonstrating remarkable rendering quality \yl{and real-time speed}, overcoming the drawbacks of NeRF~\cite{chen2024survey}. It can be used for 3D or 4D reconstruction~\cite{luiten2023dynamic, yang2023deformable3dgs}, \yl{avatar} modeling~\cite{li2023animatable,kocabas2023hugs}, etc. 
GS-based generation has also gained significant \hb{attentions} %
\cite{chen2023text, tang2023dreamgaussian, yi2023gaussiandreamer}. 
3DGS facilitates numerous applications thanks to its efficient differentiable rendering and high-fidelity \yl{rendering}. Our proposed method is built upon 3DGS, and inherits the rendering speed and the power to express \yl{detailed} appearance of 3DGS, while additionally providing \yl{full} control of the deformation of geometry.

\subsection{3D Shape Deformation \& Editing}
\noindent\textbf{Mesh Editing}
Editing a 3D model involves altering \hb{its} shape 
according to user-defined \yl{boundary conditions}.
There are \yl{many} different ways to edit a 3D model, such as Laplacian coordinates~\cite{Lipman2005LaplacianFF, SorkineHornung2005LaplacianMP, sorkine2007rigid}, Poisson \yl{equations}~\cite{yu2004mesh} and cage-based approaches~\cite{wang2020NeuralCage, zhang2020proxy}, data-driven mesh deformation~\cite{gao2019sparse,sumner2005mesh}.
These methods can be conducted in real time while preserving the geometric details. However, they can only be done on explicit \yl{mesh} representations, hindering further usage on modern %
representations \yl{that capture both geometry and appearance. }

\noindent\textbf{NeRF Editing} 
\hb{As a pioneering work for NeRF editing,} 
EditNeRF~\cite{liu2021editing} %
successfully modifies the shape and color of neural fields by including latent codes as conditioning factors. 
Furthermore, several works~\cite{wang2022clip,wang2023nerf,gao2023textdeformer,bao2023sine} utilize the CLIP model~\cite{wang2022clip} to facilitate the editing on NeRFs via text prompts or reference images.
Another stream is to use some predefined explicit proxies to support the editing, such as skeletons for \yl{humans}~\cite{peng2021neural,jiang2023avatarcraft} \yl{and} explicit geometries~\cite{Yuan2022NeRFEditingGE, yang2022neumesh, xu2022deforming, jambon2023nerfshop}, which all transfer the edits on explicit proxies to NeRF.
Besides, some 2D image manipulation (\eg inpainting, strokes) is also adopted into the NeRF editing via optimization \yl{schemes}~\cite{liu2022nerf, kobayashi2022decomposing,zhuang2023dreameditor,wang2023seal} or \yl{applied to} dynamic NeRFs~\cite{pumarola2021d} for 4D-editing~\cite{jiang20234d}. 

Although it is an innovation to introduce explicit proxies to enhance editing, these works have limited \yl{applicability} due to the high computational cost and slow rendering speed of NeRF.

\noindent\textbf{Gaussian Editing}
In \yl{contrast}, 3DGS enables a small training cost and high-quality real-time rendering via splatting rendering. 
The editing on 3DGS has also been %
explored in various fields.
PhysGaussian~\cite{xie2023physgaussian} employs discrete particle clouds from 3DGS for physically-based dynamics and photo-realistic rendering through continuum deformation~\cite{bonet1997nonlinear} of \yl{Gaussian} kernels. 
SC-GS~\cite{huang2023sc} learns sparse control points for 3D scene dynamics but faces challenges with %
\yl{large} movements and complex surface deformation. 
SuGaR~\cite{guedon2023sugar} extracts explicit meshes from 3DGS representations, but relies on simple adjustment of Gaussian parameters based on deformed meshes and struggles with large-scale deformation.
The above approaches edit the original GS by merely adjusting parameters, limiting \yl{their} effectiveness in interactive and large-scale deformation. Our observation is that 3DGS, which is based on discrete and \yl{unstructured} Gaussian kernels, needs strong topological information to guide the relationship of neighboring kernels in order to perform large-scale deformation while preserving meaningful appearance.

In this work, we pioneer the adaptation of mesh-based deformation to 3DGS by harnessing the priors of explicit representation: the surface properties like \yl{normals} of the mesh, and the gradients generated by explicit deformation methods. The full utilization of explicit mesh representation provides adequate \yl{topological} information to 3DGS, and improves both the learning and deformation of our method.

\begin{figure*}[htp]
    \includegraphics[width=0.95\linewidth]{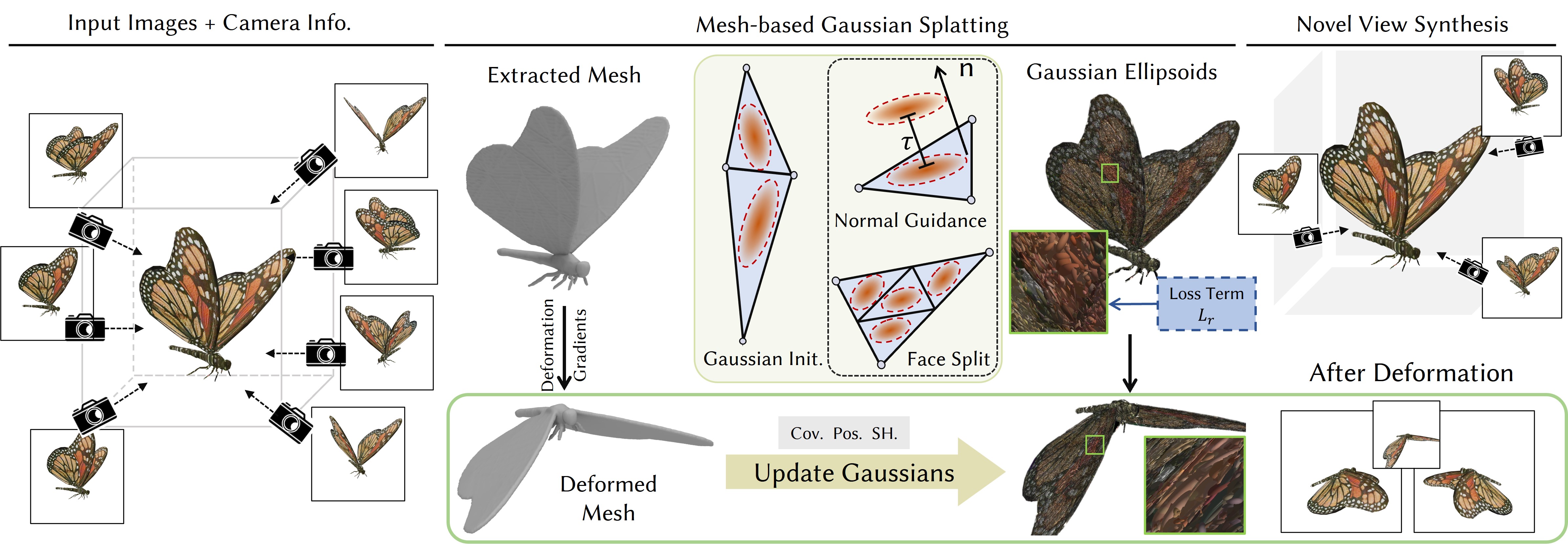}
    \caption{
    \textbf{An overview of our pipeline.} 
    \hb{Our} %
    mesh-based Gaussian \yl{splatting} representation \hb{is} specifically designed for Gaussian \hb{deformation}. %
    Given a set of calibrated images, we first reconstruct \hb{an} %
    explicit mesh using the explicit geometry prior to initialize the Gaussian.
    During the learning, the explicit mesh
    guides the Gaussian learning %
    \yl{in two strategies}
    according to the explicit mesh: a) Face Split; b) Normal Guidance.
    Also, we introduce a regularization loss $L_r$ to constrain the \yl{scale} of Gaussians to prevent the abnormally shaped Gaussians \yl{with extreme anisotropy}.
    Furthermore, the mesh-based deformation incorporates our mesh-based Gaussian \yl{splatting} representation \hb{to achieve real-time deformation on 3DGS}. %
    The deformations on \hb{the} explicit mesh (deformation gradients) from the user's controls %
    drive the parameters of Gaussians and produce the deformed 3DGS for novel view rendering.
    Overall, our pipeline not only achieves accurate and realistic rendering from novel views but also supports effortless and real-time \hb{deformation} %
    of 3DGS.
    }
    \label{fig:pipeline}
\end{figure*}

\section{Methodology}\label{sec:method}
Given a collection of multi-view images, 
we express the geometry and appearance of the scene as a Gaussian distribution \yl{bound} with an explicit mesh extracted by existing approaches (\eg NeuS2~\cite{wang2023neus2}).
Our goal is to enable real-time \yl{deformation} of 3DGS.
For realistic and interactive \yl{deformation}, our solution is to introduce an explicit mesh as the topological \yl{guidance} and use mesh-based Gaussian distribution learning to constrain the parameters and the growth process of Gaussian functions, thus ensuring the correlation between the 3D Gaussians and the geometric shape.
After \yl{Gaussian} distribution learning, 
\yl{thanks to the binding of GS and the mesh,}
the deformation gradients from the user-controlled deformed mesh are applied to the \yl{Gaussians'} parameters.
In addition, a regularization of the Gaussian shape is designed to eliminate \yl{extreme} anisotropy of the Gaussians in the \yl{deformation} process.
Hence, we incorporate them together into an interactive tool for real-time \yl{deformation} to enable photorealistic novel view rendering efficiently following the user's control.
An overview of our pipeline is presented in \yl{Figure}~\ref{fig:pipeline}.
In the following, we first preview some preliminaries, including the 3DGS representation and mesh deformation in Sec.~\ref{sec:pre}, then introduce our mesh-\yl{based} GS representation in Sec.~\ref{sec:rep} \yl{followed by} the \yl{deformation} technique in Sec.~\ref{sec:editing}.

\subsection{Preliminaries}\label{sec:pre}
\noindent\textbf{3D Gaussian Splatting}\label{pre: gs}
Each Gaussian element is defined by a center position $\mu \in \mathbb{R}^3$, a covariance matrix $\Sigma \in \mathbb{R}^7$, color $c \in \mathbb{R}^k$ (represented by spherical \yl{harmonic} coefficients for view-dependent color, where $k$ represents the degrees of freedom), and opacity $\sigma \in \mathbb{R}$. The Gaussian function $g(x)$ can be defined by the following formulation:
\begin{equation}
\label{formula:gaussian's formula}
    g(x)=e^{-\frac{1}{2}(x-\mu)^T\Sigma^{-1}(x-\mu)}
\end{equation}
where the covariance matrix $\Sigma$ can be factorized into a rotation matrix $\mathbf{R}$ expressed as a quaternion $q \in \mathbb{R}^4$ and a scaling matrix $\mathbf{S}$ represented by a 3D-vector $s \in \mathbb{R}^3$ for the differentiable optimization:
$\Sigma = \mathbf{R}\mathbf{S}\mathbf{S}^T\mathbf{R}^T$.

The rendering technique of splatting, as initially introduced in ~\cite{zwicker2001surfacesplatting}, is to project the Gaussians onto the camera image planes, \yl{which are} employed to generate novel view images. The location and covariance of the projected 2D Gaussian can be expressed as follows: $\Sigma^{\prime} = JW\Sigma W^TJ^T$ and $\mu^{\prime} = JW\mu$, \yl{which} 
involves a \yl{view} transformation $W$ and the Jacobian $J$ of the affine approximation of the projective transformation.
Specifically, the color of each Gaussian is assigned to every pixel based on the Gaussian representation described in Equ.~\ref{formula:gaussian's formula}. And the opacity controls the influence of each Gaussian.
The per-pixel color $C$ is formulated as the weighted sum of $N$ ordered Gaussians that are associated with the pixel:
$C = \sum_{i\in N}SH(d_i, c_i) \sigma_i^\prime \prod_{j=1}^{i-1} (1-\sigma^\prime_j)$,
where $SH(\cdot,\cdot)$ is the spherical harmonic function with \yl{input} direction $d_i$ and SH coefficients $c_i$, $\sigma_i^\prime$ \yl{represents} the $z$-depth ordered effective opacity: $\sigma_i^\prime = \sigma_i e^{-\frac{1}{2} ({p} - \mu_i^{\prime})^T \Sigma_i^{\prime} ({p} - \mu_i^{\prime}) }$.

\noindent\textbf{Mesh Deformation}\label{pre: deformation}
Given a triangular mesh $\mathcal{M}$ %
we can deform the mesh that satisfies the user's control (\eg sparse control points) using the \yl{mesh} deformation method ~\cite{sumner2005mesh,sorkine2007rigid,gao2019sparse}. The method~\cite{gao2019sparse} could handle large-scale mesh deformations by optimizing the deformation gradients. 
The deformation is formulated by minimizing the  energy function, which ensures the stiffness of the mesh and prevents local distortions:
\begin{equation}\label{equ:deformation energy}
    \mathcal{E}\left(\mathcal{M}^\prime\right)=\sum_{i=1}^N \sum_{j \in \mathcal{N}(i)} w_{i j}\left\|\left(\boldsymbol{v}_i^{\prime}-\boldsymbol{v}_j^{\prime}\right)-\mathbf{T}_i\left(\boldsymbol{v}_i-\boldsymbol{v}_j\right)\right\|^2
\end{equation}
where $\mathcal{N}(i)$ represents the set of 1-ring neighboring vertices of vertex $i$, $v_i$ is the spatial \yl{coordinates} of the $i^{\rm th}$ vertex on mesh $\mathcal{M}$, $v_i^\prime$ is the spatial \yl{coordinates} of the corresponding \yl{vertex} on the deformed mesh $\mathcal{M}^\prime$, 
and $w_{ij}$ is the cotangent weight ~\cite{sorkine2007rigid}, %
The transformation matrix $\mathbf{T}_i$ can be decomposed into a rotation matrix $\mathbf{\bar{R}}_i$ and a shear matrix $\mathbf{\bar{S}}_i$ by polar decomposition. Both matrices will be applied to the corresponding Gaussians during the \yl{deformation} procedure.
Note that this deformation formulation also supports data-driven \yl{deformation} if some prior exemplar 
\yl{deformed meshes} are available. The $\mathbf{T}_i$ could be optimized by blending the existed deformation gradients from the exemplar meshes.  %

\subsection{Mesh-based Gaussian Splatting}\label{sec:rep}

While the 3DGS can produce realistic \yl{rendered} images in real time, it struggles to accurately represent the details and topological structure of a 3D scene. This limitation arises from its reliance on discrete \yl{Gaussian kernels}, \yl{in particular} when it comes to \yl{deformation}.
In order to tackle these problems, we \yl{introduce our Mesh-based Gaussian Splatting}. \yl{Our} method focuses on integrating 3D Gaussian kernels with specified mesh surfaces, hence enhancing the process of Gaussian \yl{deformation}.

A reconstructed mesh $\mathcal{M}$, obtained using an existing efficient method~\cite{wang2023neus2} 
is used as \yl{an} explicit prior \yl{constraint}. It is combined with two \yl{strategies} to regulate the \yl{Gaussian} parameters and growth of Gaussian kernels, as illustrated in Figure~\ref{fig:pipeline}.
This ensures \yl{the} correlation between 3D \yl{Gaussians} and the explicit prior. 
The objective of these two \yl{strategies} is to regularize 3D Gaussian kernels while also maintaining their ability to accurately represent geometric and textural features. 
First, we initialize \yl{a} Gaussian by anchoring it precisely at the centroid of every triangular face on the  mesh \yl{surface}.
During training of \yl{mesh-based} 3DGS, different from the original 3DGS, we allow the division of Gaussians by utilizing the following formulas:
\begin{itemize}
    \item Face Split: A single triangle is subdivided into four smaller triangles over the surface \yl{by inserting a new vertex at midpoint of each triangle. The Gaussian kernels are also split in the same way.}
    \item Normal Guidance: Each Gaussian has a perpendicular movement to the surface under normal guidance. The distance of this movement $\tau$ is learnable.
\end{itemize}
The distribution of 3D \yl{Gaussians} is determined by the two \yl{strategies} mentioned above, which incorporate explicit mesh priors.
The first objective is to guarantee a sufficient number of Gaussian kernels in order to accurately represent the visual appearance of the 3D scene, following the guidance of the mesh surface.
The latter one aims to represent the fine-grained texture details of a 3D scene for \yl{Novel View Synthesis (NVS)}.
Both of them mandate the distribution of Gaussian kernels near the explicit surface beforehand. Note that the reduction \yl{operation} \yl{follows} the original 3DGS.

Thus, the barycentric coordinates $w = (w_a, w_b, w_c)$ and the offset distance $\tau$ are parameterized into additional attributes for 3DGS learning. 
The barycentric coordinates $(w_a, w_b, w_c)$ represent the weights assigned to three vertices $(\mathbf{v}_a, \mathbf{v}_b, \mathbf{v}_c)$ belonging to \yl{the nearby triangle},  $\tau \in [-0.5, 0.5]$ is the displacement along the surface normal $\mathbf{n}$. 
To summarize, the spatial position $\mu$ of the Gaussian kernel is formulated as:
\begin{equation}\label{equ:gaussian}
    \mu = (w_a \mathbf{v}_a + w_b \mathbf{v}_b + w_c \mathbf{v}_c) + \tau R \mathbf{n}
\end{equation}
where $R$ is the radius of the circumcircle \yl{of the} nearby triangle.
By utilizing the prior of explicit meshes, we employ the aforementioned \yl{strategies} (illustrated in Figure~\ref{fig:pipeline}) to regularize the density of Gaussians according to the explicit surface, and generate new Gaussian kernels, followed by Equ.~\ref{equ:gaussian}, that can be used to continue participating in the optimization.

\noindent\textbf{Regularization $L_r$}
For better \yl{visual quality} of deformed 3DGS, we introduce regularization to ensure the spatial coherence and local consistency of the Gaussians.
Since we support arbitrary \yl{deformation}, it is inevitable that local meshes \yl{can} undergo drastic changes with large-scale deformation.
It will lead to visual artifacts due to the anisotropy of 3D Gaussian when the learned Gaussian shape is large enough and covers multiple triangles on the surface.
In order to ensure plausible \yl{deformation} results, 
we employ a regularization $L_r$ that adjusts the Gaussian shape based on the size of neighboring triangles during training, which ensures that the appropriate Gaussian is learned and local continuity \yl{is preserved during deformation}.
\yl{The} regularization $L_r$ is formulated as:
\begin{equation}
    L_r=\sum_{g \in G} \max\left(\max\left(s_i\right)-\gamma R_i, 0\right)
\end{equation}
where $s_i$ is the 3D scaling vector of each \yl{Gaussian}, $R_i$ is the radius of the circumcircle of the triangle where \yl{the} Gaussian is located, and $\gamma$ is a hyper-parameter to control the influence on the size of the Gaussian from the neighboring triangles.

\subsection{Editable Gaussian with Mesh Deformation}\label{sec:editing}

By utilizing existing mesh deformation methods and taking use of the efficient differentiable rasterization of GS, it is possible to achieve real-time deformation of Gaussians based on the GS representation proposed in Sec.~\ref{sec:rep}.
To illustrate this idea, we employ the mesh deformation techniques and formulations discussed in Sec.~\ref{pre: deformation}. 
The user can manipulate 3D Gaussians using various controls, such as non-rigid deformation, translation, rotation, etc.
Equ.~\ref{equ:deformation energy} states that each vertex $v_i$ in the deformed mesh $\mathcal{M}$ is linked to a transformation matrix $T_i$, 
which represents local changes around vertex $v_i$ between the deformed mesh \yl{$\mathcal{M}^\prime$} and the original mesh \yl{$\mathcal{M}$} and can be decomposed into a rotation matrix $\mathbf{\bar{R}}_i$ and a shear matrix $\mathbf{\bar{S}}_i$ using the polar decomposition.
We can demonstrate that the Gaussian distribution remains unchanged following an affine transformation.
Thus, we can easily apply the rotation matrix $\mathbf{\bar{R}}_i$ and shear matrix $\mathbf{\bar{S}}_i$ from \yl{the} deformed mesh to its associated \yl{Gaussian} kernels, as well as the displacement of deformed mesh faces. 

For each deformed \yl{Gaussians} $g^\prime$, that \yl{is bound} with a triangle $f^\prime = (\mathbf{v}_a^\prime, \mathbf{v}_b^\prime, \mathbf{v}_c^\prime)$ with three deformed vertices.
The relative displacements $\Delta P$ and deformation gradients $T_i$ for the deformed face can be expressed using barycentric coordinates $(w_a,w_b,w_c)$:
\begin{equation}
    \begin{split}
    \Delta P &= w_a( \mathbf{v}_a^\prime -\mathbf{v}_a) +  w_b (\mathbf{v}_b^\prime - \mathbf{v}_b) + w_c (\mathbf{v}_c^\prime +\mathbf{v}_c) \\
    \mathbf{\bar{R}}_i &= w_a log(\mathbf{\bar{R}}_{\mathbf{v}_a^\prime}) + w_b log(\mathbf{\bar{R}}_{\mathbf{v}_b^\prime}) + w_c log(\mathbf{\bar{R}}_{\mathbf{v}_c^\prime})\\
    \mathbf{\bar{S}}_i &= w_a \mathbf{\bar{S}}_{\mathbf{v}_a^\prime} + w_b \mathbf{\bar{S}}_{\mathbf{v}_b^\prime} + w_c \mathbf{\bar{S}}_{\mathbf{v}_c^\prime}, T_i = exp(\mathbf{\bar{R}}_i) \mathbf{\bar{S}}_i
    \end{split}
\end{equation}
Following the above equations, we can get the transformed Gaussian kernels with position $\mu^\prime = \mu + \Delta P$ and covariance matrix $\Sigma^\prime = T_i \Sigma T_i^\top$. The deformed Gaussian kernel is 
\begin{equation}
    \label{formula:gaussian's formula new}
        g^\prime (x)=e^{-\frac{1}{2}(x-(\mu + \Delta P))^T(T_i \Sigma T_i^T)^{-1}(x-(\mu + \Delta P))}
\end{equation}
In addition, 3DGS employs spherical harmonics to express color, wherein a given Gaussian exhibits varying colors when viewed from different angles, enabling the modeling of view-dependent \yl{appearance}.
Hence, for a deformed Gaussian kernel $g^\prime$, 
it is necessary to adjust the orientation of spherical harmonics~\cite{guedon2023sugar} by applying the inverse of the local rotation matrix $exp(\mathbf{\bar{R}}_i)$ from the deformed mesh to the view direction $d$, \ie $SH(exp(\mathbf{\bar{R}}_i)^T d, c_i)$.
In conclusion, \yl{our} mesh-based GS representation allows flexible manipulation of Gaussians through mesh deformation, and the high-fidelity rendering results in novel views.

\noindent\textbf{Real-Time Interactive Tool}
By utilizing the above approach, we integrate them into an interactive deformation tool that allows for the real-time deformation of 3DGS via the user's controls. Users can utilize the triangle mesh as a proxy to accomplish large-scale real-time deformation of 3DGS, as well as high-fidelity rendering results.

\newcommand{\zoomin}[9]{ %
\begin{tikzpicture}[spy using outlines={rectangle,#9,magnification=#8,size=#6}]   
	\node[anchor=south west,inner sep=0]  {\includegraphics[width=#7]{#1}};
	\spy on (#2, #3) in node at (#4,#5);
\end{tikzpicture}
}

\newcommand{\zoominhere}[9]{ %
\begin{tikzpicture}[spy using outlines={rectangle,#9,magnification=#8,size=#6}]   
	\node[anchor=south west,inner sep=0]  {\includegraphics[width=#7,trim={8cm 8cm 8cm 8cm},clip]{#1}};
	\spy on (#2, #3) in node at (#4,#5);
\end{tikzpicture}
}

\newcommand{\arrowin}[4]{
\begin{tikzpicture}
            \node[anchor=south west,inner sep=0] (image) at (0,0) {\includegraphics[width=#3,trim={#2 #2 #2 #2},clip]{#1}};
            \begin{scope}[x={(image.south east)},y={(image.north west)}]
                \draw[#4,ultra thick,->] (0.28,0.28) -- (0.43,0.43);
            \end{scope}
        \end{tikzpicture}
}

\newcommand{\zoomintrans}[9]{
\begin{tikzpicture}[spy using outlines={rectangle,#9,magnification=#8,size=#6}]   
	\node[anchor=north west,inner sep=0]  {\includegraphics[width=#7,trim={-8cm 0cm 8cm 0cm},clip]{#1}};
	\spy on (#2, #3) in node at (#4,#5);
\end{tikzpicture}
}

\newcommand{\zoomindeer}[9]{ %
\begin{tikzpicture}[spy using outlines={rectangle,#9,magnification=#8,size=#6}]   
	\node[anchor=north east,inner sep=0]  {\includegraphics[width=#7,trim={7cm 7cm 7cm 7cm},clip]{#1}};
	\spy on (#2, #3) in node at (#4,#5);
\end{tikzpicture}
}

\begin{figure*}[!h]
	\newlength\mytmplen
	\setlength\mytmplen{.102\linewidth}
	\setlength{\tabcolsep}{1pt}
	\renewcommand{\arraystretch}{0.5}
	\centering
	\begin{tabular}{c|cccccccc}
		Input \& Edit&\multicolumn{2}{c}{NeRF-Editing} & \multicolumn{2}{c}{Baseline} & \multicolumn{2}{c}{SuGaR} & \multicolumn{2}{c}{Ours} \\
          \includegraphics[width=\mytmplen,trim={5cm 5cm 5cm 5cm},clip]{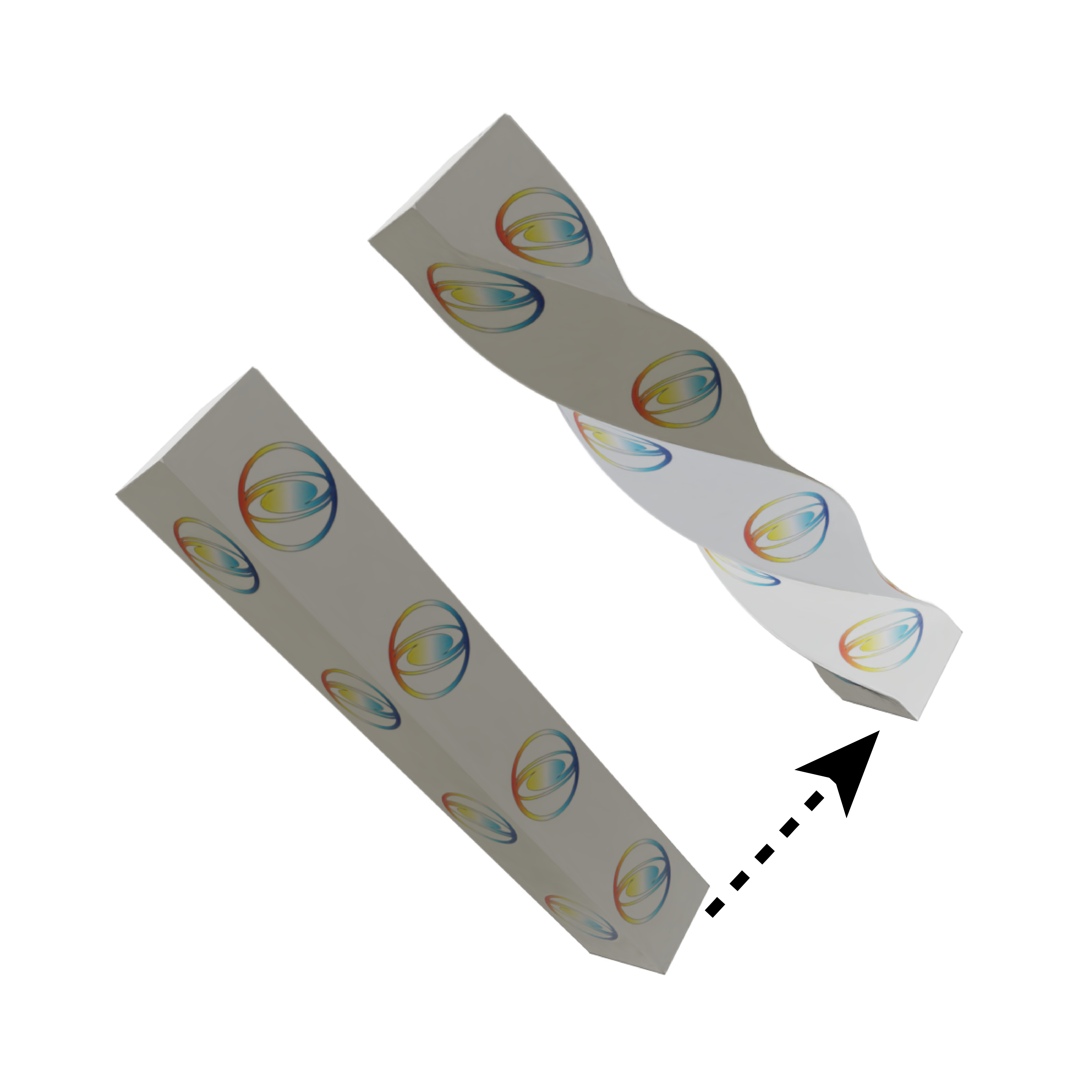} &
        \arrowin{comparison/square/00011-NE.png}{8cm}{\mytmplen}{red}&
        \arrowin{comparison/square/00016-NE.png}{8cm}{\mytmplen}{cyan}&
        \arrowin{comparison/square/00011-base.png}{8cm}{\mytmplen}{red} &
        \arrowin{comparison/square/00016-base.png}{8cm}{\mytmplen}{cyan} &
        \zoominhere{comparison/square/00011-sugar.png}{1.1}{0.7}{0.4cm}{0.4cm}{0.8cm}{\mytmplen}{4}{red} &
        \zoominhere{comparison/square/00016-sugar.png}{1.4}{0.25}{0.4cm}{0.4cm}{0.8cm}{\mytmplen}{4}{cyan} &
        \zoominhere{comparison/square/00011-ours.png}{1.1}{0.7}{0.4cm}{0.4cm}{0.8cm}{\mytmplen}{4}{red} &
        \zoominhere{comparison/square/00016-ours.png}{1.4}{0.25}{0.4cm}{0.4cm}{0.8cm}{\mytmplen}{4}{cyan}
        \\
		\includegraphics[width=\mytmplen]{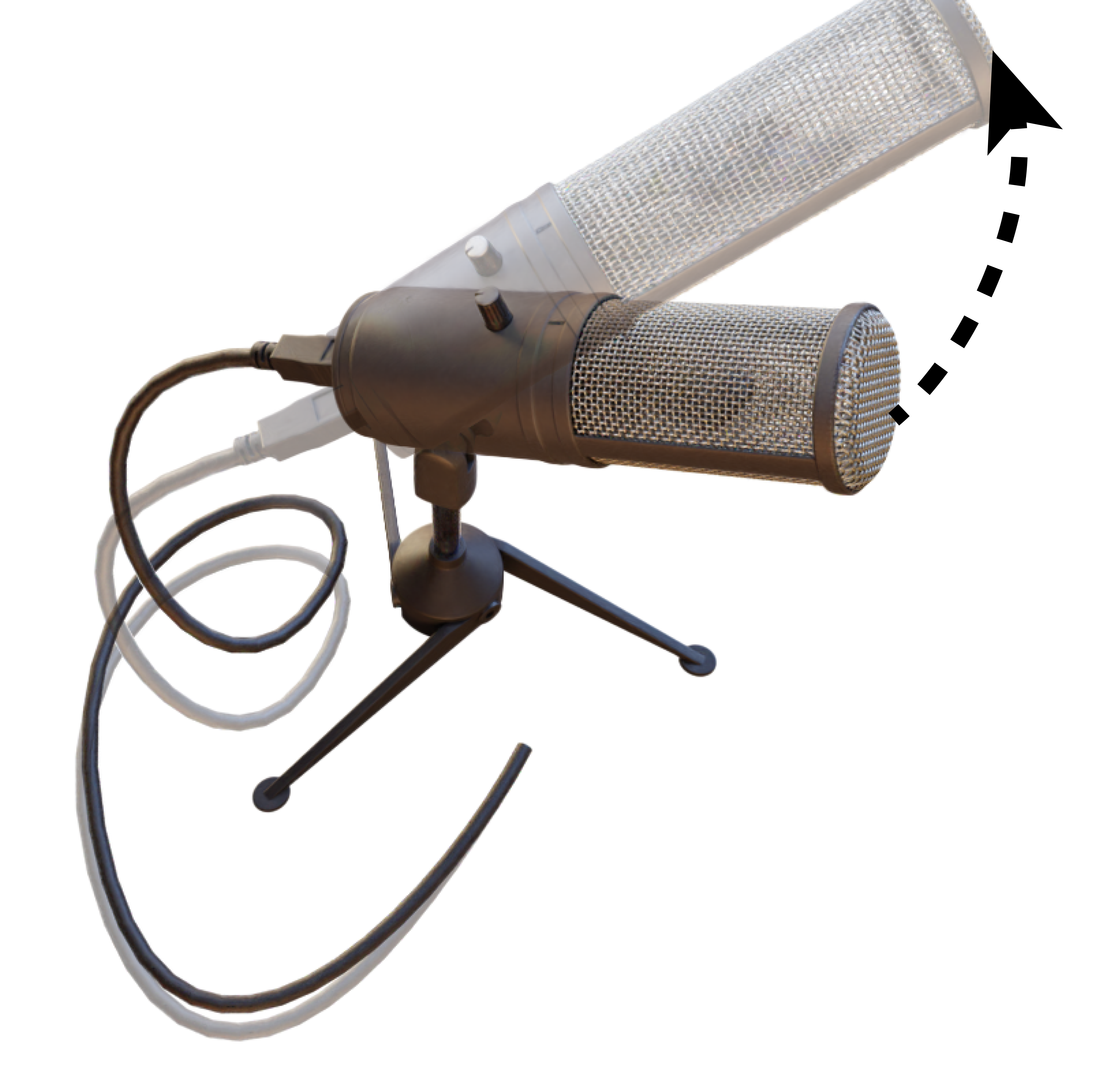} &
        \zoomin{comparison/mic/00064-nerfediting.png}{0.4}{1.3}{0.55cm}{0.55cm}{1cm}{\mytmplen}{3}{red} &
        \zoomin{comparison/mic/00070-nerfediting.png}{1.55}{1.6}{0.55cm}{0.55cm}{1cm}{\mytmplen}{3}{cyan} &
        \zoomin{comparison/mic/00064-base.png}{0.4}{1.3}{0.55cm}{0.55cm}{1cm}{\mytmplen}{3}{red} &
        \zoomin{comparison/mic/00070-base.png}{1.55}{1.6}{0.55cm}{0.55cm}{1cm}{\mytmplen}{3}{cyan} &
		\zoomin{comparison/mic/00064-sugar.png}{0.4}{1.3}{0.55cm}{0.55cm}{1cm}{\mytmplen}{3}{red} &
        \zoomin{comparison/mic/00070-sugar.png}{1.55}{1.6}{0.55cm}{0.55cm}{1cm}{\mytmplen}{3}{cyan} &
        \zoomin{comparison/mic/00064-ours.png}{0.4}{1.3}{0.55cm}{0.55cm}{1cm}{\mytmplen}{3}{red} &
        \zoomin{comparison/mic/00070-ours.png}{1.55}{1.6}{0.55cm}{0.55cm}{1cm}{\mytmplen}{3}{cyan}
        \\
		\includegraphics[width=\mytmplen,trim={5cm 5cm 5cm 5cm},clip]{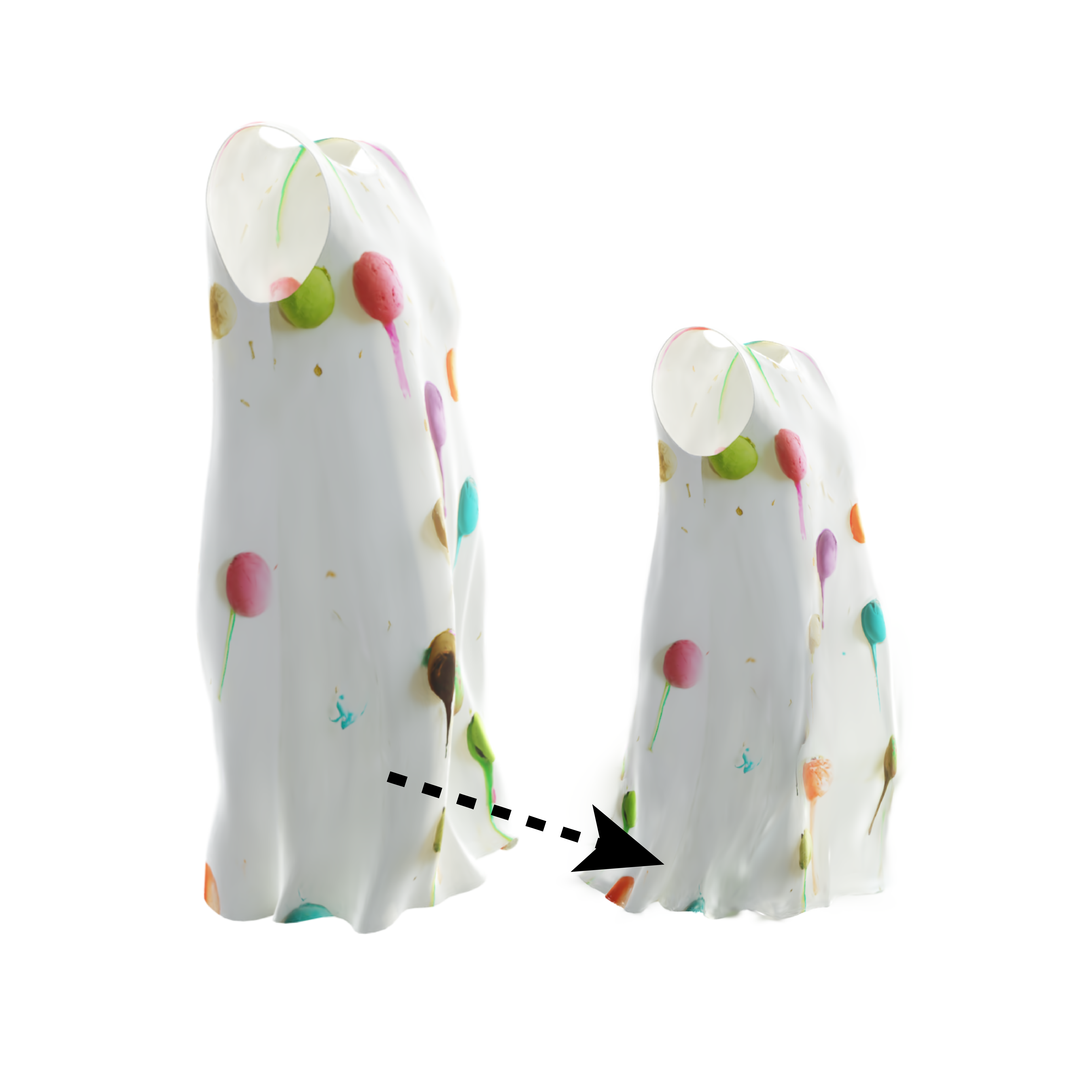} &
        \zoomintrans{comparison/dress/00017-NE.png}{1.3}{-1.6}{0.55cm}{-0.55cm}{1cm}{\mytmplen}{3}{red} &
        \zoomintrans{comparison/dress/00105-NE.png}{1.5}{-1.45}{0.55cm}{-0.55cm}{1cm}{\mytmplen}{3}{cyan} &
        \zoomintrans{comparison/dress/00017-base.png}{1.3}{-1.6}{0.55cm}{-0.55cm}{1cm}{\mytmplen}{3}{red} &
        \zoomintrans{comparison/dress/00105-base.png}{1.5}{-1.45}{0.55cm}{-0.55cm}{1cm}{\mytmplen}{3}{cyan} &
		\zoomintrans{comparison/dress/00017-sugar.png}{1.3}{-1.6}{0.55cm}{-0.55cm}{1cm}{\mytmplen}{3}{red} &
        \zoomintrans{comparison/dress/00105-sugar.png}{1.5}{-1.45}{0.55cm}{-0.55cm}{1cm}{\mytmplen}{3}{cyan} &
        \zoomintrans{comparison/dress/00017-ours.png}{1.3}{-1.6}{0.55cm}{-0.55cm}{1cm}{\mytmplen}{3}{red} &
        \zoomintrans{comparison/dress/00105-ours.png}{1.5}{-1.45}{0.55cm}{-0.55cm}{1cm}{\mytmplen}{3}{cyan}
        \\
        \includegraphics[width=\mytmplen,trim={1cm 1cm 1cm 1cm},clip]{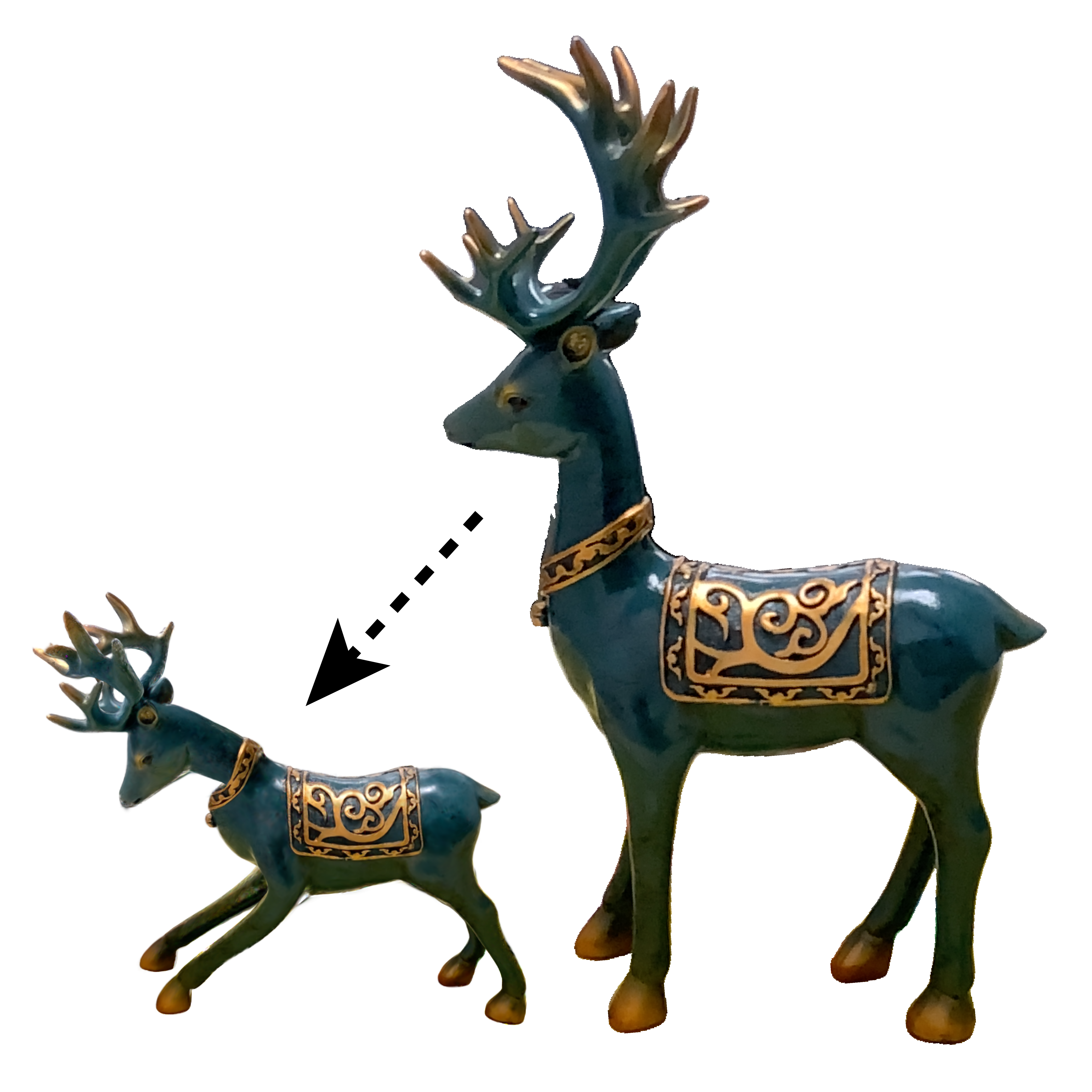} &
        \zoomindeer{comparison/deer/00109-NE-01.png}{-1.45}{-0.8}{-0.4cm}{-0.4cm}{1cm}{\mytmplen}{3}{red}&
        \zoomindeer{comparison/deer/00148-NE-01.png}{-1.0}{-0.9}{-0.4cm}{-0.4cm}{0.8cm}{\mytmplen}{4}{cyan}&
        \zoomindeer{comparison/deer/00109-base-01.png}{-1.45}{-0.8}{-0.4cm}{-0.4cm}{1cm}{\mytmplen}{3}{red}&
        \zoomindeer{comparison/deer/00148-base-01.png}{-1.0}{-0.9}{-0.4cm}{-0.4cm}{0.8cm}{\mytmplen}{4}{cyan}&
        \zoomindeer{comparison/deer/00109-sugar-01.png}{-1.45}{-0.8}{-0.4cm}{-0.4cm}{1cm}{\mytmplen}{3}{red}&
        \zoomindeer{comparison/deer/00148-sugar-01.png}{-1.0}{-0.9}{-0.4cm}{-0.4cm}{0.8cm}{\mytmplen}{4}{cyan}&
        \zoomindeer{comparison/deer/00109-ours-01.png}{-1.45}{-0.8}{-0.4cm}{-0.4cm}{1cm}{\mytmplen}{3}{red}&
        \zoomindeer{comparison/deer/00148-ours-01.png}{-1.0}{-0.9}{-0.4cm}{-0.4cm}{0.8cm}{\mytmplen}{4}{cyan}
	\end{tabular}
	\caption{
		\label{fig:comparisons}
    \textbf{Comparison with the alternative methods.} We show comparisons of ours to previous methods and the editing results from the novel views. There are \textsc{Mic} from NeRF-Synthetic, \textsc{Cubiod}, \textsc{Dress} and \textsc{Deer} captured by ourselves. We have highlighted the difference with different color boxes for different views. From the results, we can see that our method successfully preserves the high-frequency details after large-scale deformation.
	}
\end{figure*}

\newcommand{\picframebox}[3]{ %
\framebox{\includegraphics[width=#2,trim={#3 #3 #3 #3},clip]{#1}}
}

\newcommand{\picframeboxa}[4]{ %
\framebox{\includegraphics[width=#2,trim={#3 #4 #3 #4},clip]{#1}}
}

\begin{figure*}[!h]
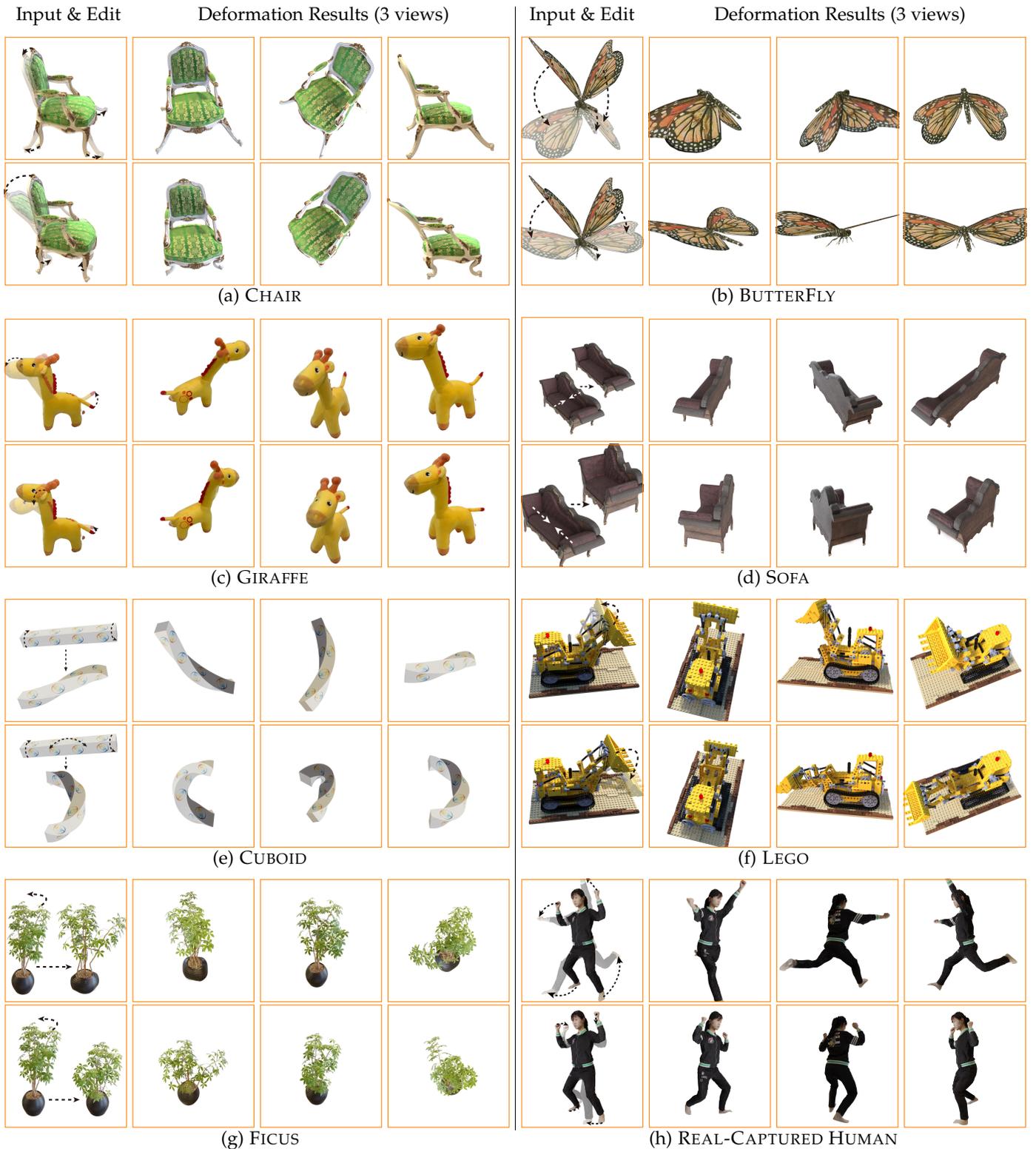

	\setlength\mytmplen{.121\linewidth}
	\setlength{\tabcolsep}{0.1pt}
        \renewcommand{\arraystretch}{0.5}
        \captionsetup[sub]{font=footnotesize,textfont=sf,labelfont=sf}
        \setlength{\fboxsep}{-0.02cm}
	\centering
	\begin{tabular}{cccc|cccc}
    Input \& Edit & \multicolumn{3}{c|}{Deformation Results (3 views)} & Input \& Edit & \multicolumn{3}{c}{Deformation Results (3 views)} \\
    \multicolumn{4}{c|}{} & \multicolumn{4}{c}{}\\
     \picframebox{moreeditresults1/chairs/00001-edit1-0-1-01.png}{\mytmplen}{1cm} &
     \picframebox{moreeditresults1/chairs/00012-d1.png}{\mytmplen}{1cm} &
     \picframebox{moreeditresults1/chairs/00045-d1.png}{\mytmplen}{1cm} &
     \picframebox{moreeditresults1/chairs/00092-d1.png}{\mytmplen}{1cm} &
     \hspace{0.1pt}\picframebox{moreeditresults1/butterfly/00000-edit1-0-01.png}{\mytmplen}{1cm} &
     \picframebox{moreeditresults1/butterfly/00009-d1.png}{\mytmplen}{1cm} &
     \picframebox{moreeditresults1/butterfly/00024-d1.png}{\mytmplen}{1cm} &
     \picframebox{moreeditresults1/butterfly/00030-d1.png}{\mytmplen}{1cm} 
     \\
      \picframebox{moreeditresults1/chairs/00001-edit2-0-1-01.png}{\mytmplen}{1cm} &
     \picframebox{moreeditresults1/chairs/00012-d2.png}{\mytmplen}{1cm} &
     \picframebox{moreeditresults1/chairs/00045-d2.png}{\mytmplen}{1cm} &
     \picframebox{moreeditresults1/chairs/00092-d2.png}{\mytmplen}{1cm} &
     \hspace{0.1pt}\picframebox{moreeditresults1/butterfly/00000-edit2-0-1-01.png}{\mytmplen}{1cm} &
     \picframebox{moreeditresults1/butterfly/00009-d2.png}{\mytmplen}{1cm} &
     \picframebox{moreeditresults1/butterfly/00024-d2.png}{\mytmplen}{1cm} &
     \picframebox{moreeditresults1/butterfly/00030-d2.png}{\mytmplen}{1cm}
     \\
    \multicolumn{4}{c|}{(a) \textsc{Chair}} & \multicolumn{4}{c}{(b) \textsc{ButterFly}} \\
     \multicolumn{4}{c|}{} & \multicolumn{4}{c}{}\\
     \picframebox{moreeditresults1/giraffe/reshape_1440/00040-edit1-0-1-01.png}{\mytmplen}{1cm} &
     \picframebox{moreeditresults1/giraffe/reshape_1440/00022-d6.png}{\mytmplen}{1cm} &
     \picframebox{moreeditresults1/giraffe/reshape_1440/00090-d6.png}{\mytmplen}{1cm} &
     \picframebox{moreeditresults1/giraffe/reshape_1440/00086-d6.png}{\mytmplen}{1cm} &
     \hspace{0.1pt}\picframebox{moreeditresults1/sofa/00019-edit1-0-1.png}{\mytmplen}{1cm} &
     \picframebox{moreeditresults1/sofa/00034-d1.png}{\mytmplen}{1cm} &
     \picframebox{moreeditresults1/sofa/00107-d1.png}{\mytmplen}{1cm} &
     \picframebox{moreeditresults1/sofa/00211-d1.png}{\mytmplen}{1cm}
     \\
     \picframebox{moreeditresults1/giraffe/reshape_1440/00040-edit2-0-1.png}{\mytmplen}{1cm} &
     \picframebox{moreeditresults1/giraffe/reshape_1440/00022-d8.png}{\mytmplen}{1cm} &
     \picframebox{moreeditresults1/giraffe/reshape_1440/00090-d8.png}{\mytmplen}{1cm} &
     \picframebox{moreeditresults1/giraffe/reshape_1440/00086-d8.png}{\mytmplen}{1cm} &
     \hspace{0.1pt}\picframebox{moreeditresults1/sofa/00019-edit2-0-1-01.png}{\mytmplen}{1cm} &
     \picframebox{moreeditresults1/sofa/00034-d2.png}{\mytmplen}{1cm} &
     \picframebox{moreeditresults1/sofa/00107-d2.png}{\mytmplen}{1cm} &
     \picframebox{moreeditresults1/sofa/00211-d2.png}{\mytmplen}{1cm}
     \\
    \multicolumn{4}{c|}{(c) \textsc{Giraffe}} & \multicolumn{4}{c}{(d) \textsc{Sofa}} \\
     \multicolumn{4}{c|}{} & \multicolumn{4}{c}{}\\
     \picframebox{moreeditresults1/newsquare/00009-edit1-0-1-01.png}{\mytmplen}{1cm} &
     \picframebox{moreeditresults1/newsquare/00012-d1.png}{\mytmplen}{1cm} &
     \picframebox{moreeditresults1/newsquare/00019-d1.png}{\mytmplen}{1cm} &
     \picframebox{moreeditresults1/newsquare/00103-d1.png}{\mytmplen}{1cm} &
     \hspace{0.1pt}\picframebox{moreeditresults1/lego/00005-edit1-0-1.png}{\mytmplen}{1cm} &
     \picframebox{moreeditresults1/lego/00012-d1.png}{\mytmplen}{1cm} &
     \picframebox{moreeditresults1/lego/00025-d1.png}{\mytmplen}{1cm} &
     \picframebox{moreeditresults1/lego/00049-d1.png}{\mytmplen}{1cm}
     \\
     \picframebox{moreeditresults1/newsquare/00009-edit2-0-1-01.png}{\mytmplen}{1cm} &
     \picframebox{moreeditresults1/newsquare/00012-d2.png}{\mytmplen}{1cm} &
     \picframebox{moreeditresults1/newsquare/00019-d2.png}{\mytmplen}{1cm} &
     \picframebox{moreeditresults1/newsquare/00103-d2.png}{\mytmplen}{1cm} &
     \hspace{0.1pt}\picframebox{moreeditresults1/lego/00005-edit2-0-1-01.png}{\mytmplen}{1cm} &
     \picframebox{moreeditresults1/lego/00012-d2.png}{\mytmplen}{1cm} &
     \picframebox{moreeditresults1/lego/00025-d2.png}{\mytmplen}{1cm} &
     \picframebox{moreeditresults1/lego/00049-d2.png}{\mytmplen}{1cm}
     \\
    \multicolumn{4}{c|}{(e) \textsc{Cuboid}} & \multicolumn{4}{c}{(f) \textsc{Lego}} \\
     \multicolumn{4}{c|}{} & \multicolumn{4}{c}{}\\
     \picframeboxa{moreeditresults1/ficus/00045-edit1-0-1-01.png}{\mytmplen}{0.5cm}{0.5cm} &
     \picframeboxa{moreeditresults1/ficus/00013-d1.png}{\mytmplen}{0.5cm}{0.5cm} &
     \picframeboxa{moreeditresults1/ficus/00046-d1.png}{\mytmplen}{0.5cm}{0.5cm} &
     \picframeboxa{moreeditresults1/ficus/00053-d1.png}{\mytmplen}{0.5cm}{0.5cm} &
     \hspace{0.1pt}\picframebox{moreeditresults1/human/00270-edit2-01.png}{\mytmplen}{1cm} &
     \picframebox{moreeditresults1/human/00043-d2.png}{\mytmplen}{1cm} &
     \picframebox{moreeditresults1/human/00123-d2.png}{\mytmplen}{1cm} &
     \picframebox{moreeditresults1/human/00247-d2.png}{\mytmplen}{1cm}
     \\
     \picframeboxa{moreeditresults1/ficus/00045-edit2-0-1-01-01.png}{\mytmplen}{0.5cm}{0.5cm} &
     \picframeboxa{moreeditresults1/ficus/00013-d2.png}{\mytmplen}{0.5cm}{0.5cm} &
     \picframeboxa{moreeditresults1/ficus/00046-d2.png}{\mytmplen}{0.5cm}{0.5cm} &
     \picframeboxa{moreeditresults1/ficus/00053-d2.png}{\mytmplen}{0.5cm}{0.5cm} &
     \hspace{0.1pt}\picframebox{moreeditresults1/human/00270-edit3-0-01.png}{\mytmplen}{1cm} &
     \picframebox{moreeditresults1/human/00043-d3.png}{\mytmplen}{1cm} &
     \picframebox{moreeditresults1/human/00123-d3.png}{\mytmplen}{1cm} &
     \picframebox{moreeditresults1/human/00247-d3.png}{\mytmplen}{1cm}
     \\
    \multicolumn{4}{c}{(g) \textsc{Ficus}} & \multicolumn{4}{c}{(h) \textsc{Real-Captured Human}}
    \end{tabular}
    \vspace{-.2cm} %
    \caption{
        \label{fig:moreresults}
       \textbf{More 3D GS deformation results.} 
       It illustrates our proposed methods of synthesizing three novel views after making modifications using 8 examples, including \textsc{ButterFly}, \textsc{Sofa}, \textsc{Cuboid} from the SketchFab~\cite{Sketchfab}, \textsc{Giraffe} captured by ourself, \textsc{Lego}, \textsc{Ficus}, \textsc{Chair} from NeRF-Synthetic dataset~\cite{pumarola2021d}, \textsc{Real-Captured Human} from THuman3.0 Dataset~\cite{deepcloth_su2022}. Each example consists of 2 edits. 
       It is clearly shown that our results are more realistic and high-fidelity from novel view rendering, as well as the various deformations.
    }
\end{figure*}

\newcommand{\zoomincrop}[9]{ %
\begin{tikzpicture}[spy using outlines={rectangle,#9,magnification=2.5,size=#6}]   
	\node[anchor=south west,inner sep=0]  {\includegraphics[width=#7,trim={#8 #8 #8 #8},clip]{#1}};
	\spy on (#2, #3) in node at (#4,#5);
\end{tikzpicture}
}

\newcommand{\zoomincropsn}[9]{ %
\begin{tikzpicture}[spy using outlines={rectangle,#9,magnification=3,size=#6}]   
	\node[anchor=north west,inner sep=0]  {\includegraphics[width=#7,trim={#8 #8 #8 #8},clip]{#1}};
	\spy on (#2, #3) in node at (#4,#5);
\end{tikzpicture}
}

\begin{figure*}[!h]
	\setlength\mytmplen{.135\linewidth}
	\setlength{\tabcolsep}{1pt}
	\renewcommand{\arraystretch}{0.5}
	\centering
	\begin{tabular}{c|cccccc}
		Input \& Edit&\multicolumn{2}{c}{w/o Face Split} & \multicolumn{2}{c}{w/o $L_r$} & \multicolumn{2}{c}{Ours} \\
		\includegraphics[width=\mytmplen,trim={3cm 5cm 2cm 5cm},clip]{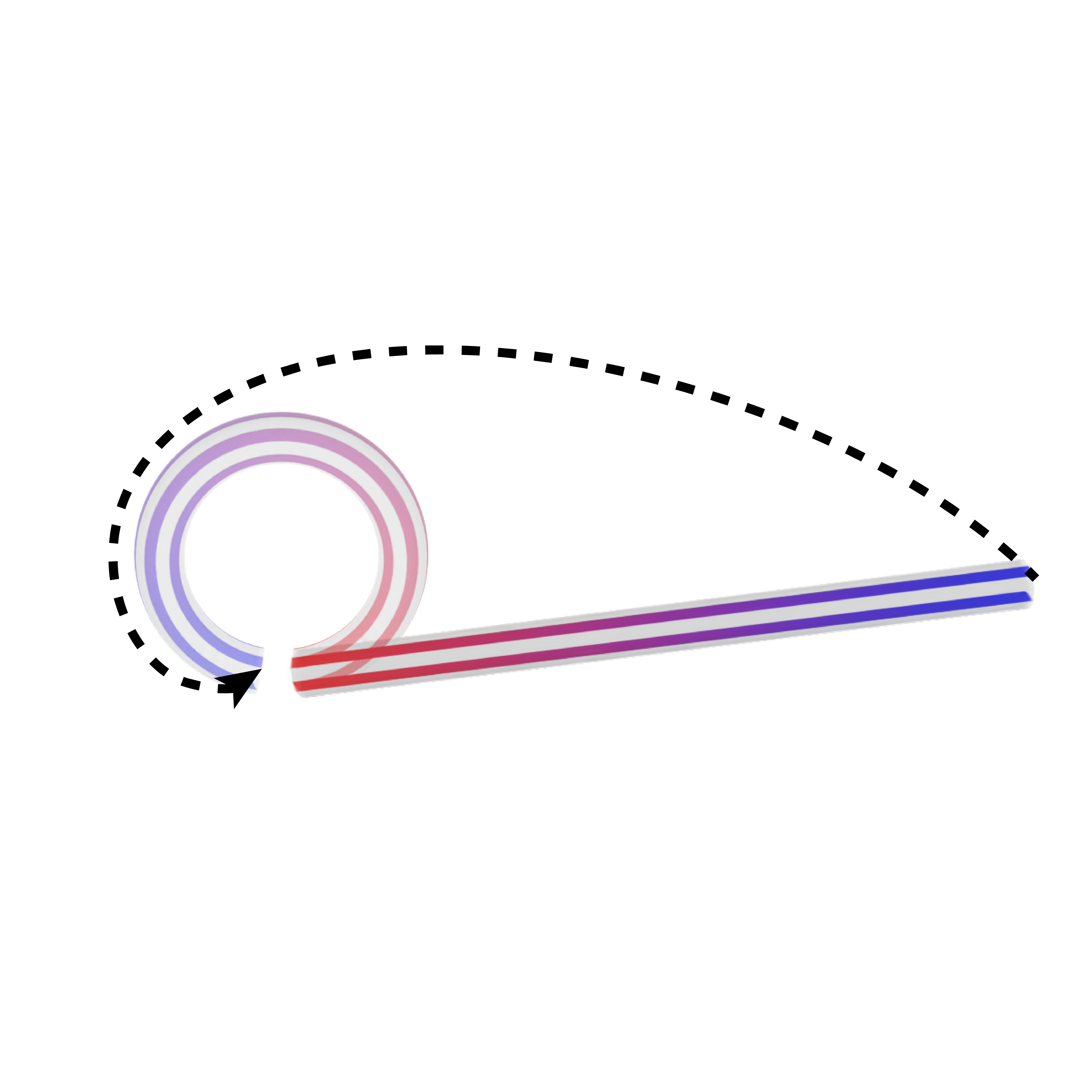} &
        \zoomincrop{ablation/loss-sub/cylinder/00283-wosub.png}{0.5}{1.3}{0.55cm}{0.55cm}{1cm}{\mytmplen}{14cm}{red} &
        \zoomincrop{ablation/loss-sub/cylinder/00285-wosub.png}{1.9}{1.3}{0.55cm}{0.55cm}{1cm}{\mytmplen}{14cm}{cyan} &
        \zoomincrop{ablation/loss-sub/cylinder/00283-woloss.png}{0.5}{1.3}{0.55cm}{0.55cm}{1cm}{\mytmplen}{14cm}{red} &
        \zoomincrop{ablation/loss-sub/cylinder/00285-woloss.png}{1.9}{1.3}{0.55cm}{0.55cm}{1cm}{\mytmplen}{14cm}{cyan} &
        \zoomincrop{ablation/loss-sub/cylinder/00283-ours.png}{0.5}{1.3}{0.55cm}{0.55cm}{1cm}{\mytmplen}{14cm}{red} &
        \zoomincrop{ablation/loss-sub/cylinder/00285-ours.png}{1.9}{1.3}{0.55cm}{0.55cm}{1cm}{\mytmplen}{14cm}{cyan}\\
		\includegraphics[width=\mytmplen,trim={5cm 5cm 5cm 5cm},clip]{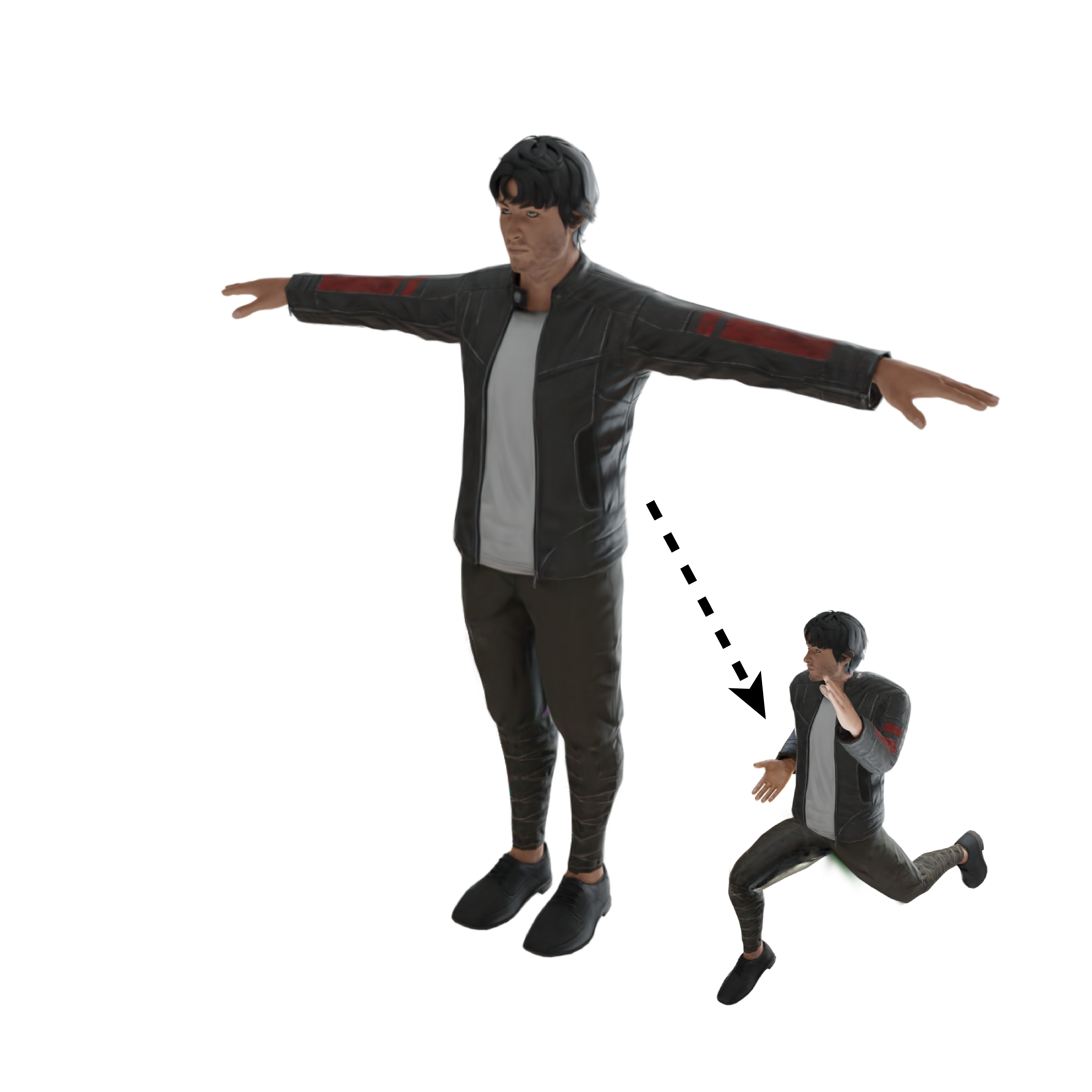} &
		\zoomincrop{ablation/loss-sub/human/00055-wosub.png}{0.7}{1.65}{0.55cm}{0.55cm}{1cm}{\mytmplen}{5cm}{red}&
		\zoomincropsn{ablation/loss-sub/human/00068-wosub.png}{1.5}{-1.}{0.55cm}{-0.6cm}{1cm}{\mytmplen}{5cm}{cyan}&
		\zoomincrop{ablation/loss-sub/human/00055-woloss.png}{0.7}{1.65}{0.55cm}{0.55cm}{1cm}{\mytmplen}{5cm}{red}&
		\zoomincropsn{ablation/loss-sub/human/00068-woloss.png}{1.5}{-1.}{0.55cm}{-0.6cm}{1cm}{\mytmplen}{5cm}{cyan}&
        \zoomincrop{ablation/loss-sub/human/00055-ours.png}{0.7}{1.65}{0.55cm}{0.55cm}{1cm}{\mytmplen}{5cm}{red}&
		\zoomincropsn{ablation/loss-sub/human/00068-ours.png}{1.5}{-1.}{0.55cm}{-0.6cm}{1cm}{\mytmplen}{5cm}{cyan}

	\end{tabular}
    \caption{
        \textbf{Ablations on Face Split operation and regularization $L_r$.} We perform the qualitative comparison on our two ablated versions: w/o Face Split and w/o $L_r$. The two cases illustrate that our full method can achieve the best results when large scale deformation appears. Especially for $L_r$, it becomes more evident that there are a greater number of Gaussians with irrational shapes when $L_r$ is not present. Disabling Face Split may result in blurry artifacts visually.
    }
    \label{fig:ablation_sub_loss}
\end{figure*}

\newcommand{\zoomincropthree}[9]{ %
\begin{tikzpicture}[spy using outlines={rectangle,#9,magnification=3,size=#6}]   
	\node[anchor=south west,inner sep=0]  {\includegraphics[width=#7,trim={#8 #8 #8 #8},clip]{#1}};
	\spy on (#2, #3) in node at (#4,#5);
\end{tikzpicture}
}

\newcommand{\zoomincroptwothree}[9]{ %
\begin{tikzpicture}[spy using outlines={rectangle,#9,magnification=2.5,size=#6}]   
	\node[anchor=south west,inner sep=0]  {\includegraphics[width=#7,trim={#8 #8 #8 #8},clip]{#1}};
	\spy on (#2, #3) in node at (#4,#5);
\end{tikzpicture}
}

\newcommand{\zoomincropsnthree}[9]{ %
\begin{tikzpicture}[spy using outlines={rectangle,#9,magnification=3,size=#6}]   
	\node[anchor=north west,inner sep=0]  {\includegraphics[width=#7,trim={#8 #8 #8 #8},clip]{#1}};
	\spy on (#2, #3) in node at (#4,#5);
\end{tikzpicture}
}

\newcommand{\zoomincropnethree}[9]{ %
\begin{tikzpicture}[spy using outlines={rectangle,#9,magnification=3,size=#6}]   
	\node[anchor=north east,inner sep=0]  {\includegraphics[width=#7,trim={#8 #8 #8 #8},clip]{#1}};
	\spy on (#2, #3) in node at (#4,#5);
\end{tikzpicture}
}

\begin{figure*}[!h]
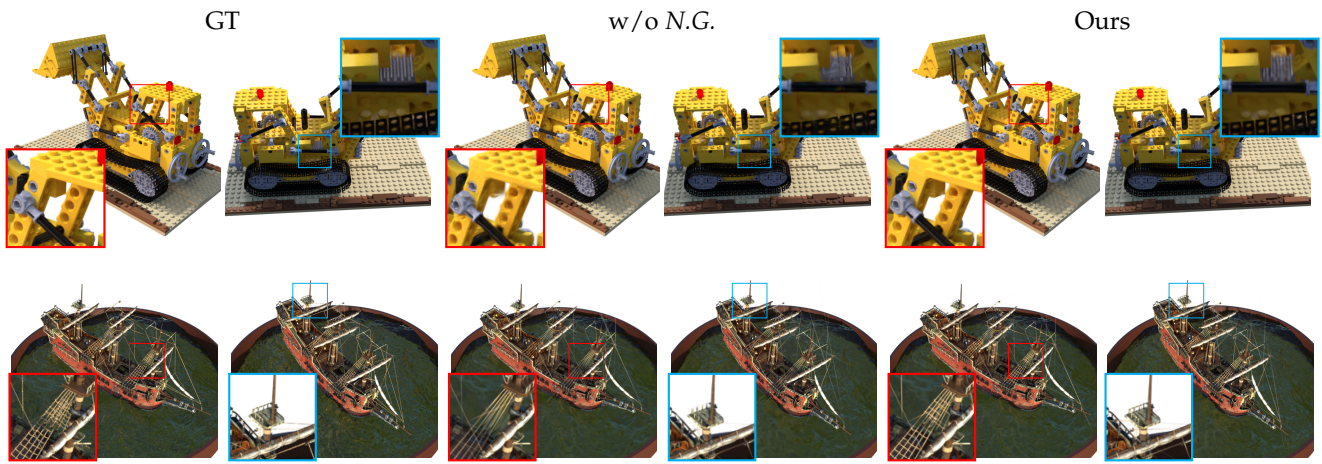

	\setlength\mytmplen{.15\linewidth}
	\setlength{\tabcolsep}{1pt}
	\renewcommand{\arraystretch}{0.5}
	\centering
 \vspace{-.1cm}
	\begin{tabular}{cccccc}
		\multicolumn{2}{c}{GT}&\multicolumn{2}{c}{w/o \textit{N.G.}} & \multicolumn{2}{c}{Ours} \\
		\zoomincroptwothree{ablation/normal/lego/00015-gt.png}{1.8}{1.8}{0.55cm}{0.55cm}{1.3cm}{\mytmplen}{3cm}{red}&
		\zoomincropnethree{ablation/normal/lego/00124-gt.png}{-1.55}{-1.45}{-0.55cm}{-0.6cm}{1.3cm}{\mytmplen}{3cm}{cyan}&
		\zoomincroptwothree{ablation/normal/lego/00015-nonormal.png}{1.8}{1.8}{0.55cm}{0.55cm}{1.3cm}{\mytmplen}{3cm}{red}&
		\zoomincropnethree{ablation/normal/lego/00124-nonormal.png}{-1.55}{-1.45}{-0.55cm}{-0.6cm}{1.3cm}{\mytmplen}{3cm}{cyan}&
        \zoomincroptwothree{ablation/normal/lego/00015-ours.png}{1.8}{1.8}{0.55cm}{0.55cm}{1.3cm}{\mytmplen}{3cm}{red}&
		\zoomincropnethree{ablation/normal/lego/00124-ours.png}{-1.55}{-1.45}{-0.55cm}{-0.6cm}{1.3cm}{\mytmplen}{3cm}{cyan}\\
        \zoomincroptwothree{ablation/normal/ship/00002-gt.png}{1.8}{1.3}{0.55cm}{0.55cm}{1.15cm}{\mytmplen}{3cm}{red}&
		\zoomincroptwothree{ablation/normal/ship/00051-gt.png}{1.05}{2.1}{0.55cm}{0.55cm}{1.15cm}{\mytmplen}{3cm}{cyan}&
		\zoomincroptwothree{ablation/normal/ship/00002-nonormal.png}{1.8}{1.3}{0.55cm}{0.55cm}{1.15cm}{\mytmplen}{3cm}{red}&
		\zoomincroptwothree{ablation/normal/ship/00051-nonormal.png}{1.05}{2.1}{0.55cm}{0.55cm}{1.15cm}{\mytmplen}{3cm}{cyan}&
        \zoomincroptwothree{ablation/normal/ship/00002-ours.png}{1.8}{1.3}{0.55cm}{0.55cm}{1.15cm}{\mytmplen}{3cm}{red}&
		\zoomincroptwothree{ablation/normal/ship/00051-ours.png}{1.05}{2.1}{0.55cm}{0.55cm}{1.15cm}{\mytmplen}{3cm}{cyan}
	\end{tabular}
    \caption{
        \textbf{Ablation of Normal Guidance on Novel View Synthesis.}
        we perform the qualitative evaluation on Normal Guidance on Novel View Synthesis task. It is more clear that Normal Guidance can enhance the high-frequency details and complex structures significantly from novel view renderings. The differences are highlighted with different colored boxes for different views  (\eg the highlighted detailed structures of the Ship and Lego examples).
    }
    \label{fig:ablation_normal}
\end{figure*}

\section{Experiments \& Evaluations}\label{sec:exps}

In this section, we conduct a series of qualitative and quantitative experiments to evaluate the effectiveness of our approach, including comparing our method to existing techniques, \yl{deformation} results on both synthetic and real-captured datasets, and ablation studies to analyze the impact of our main design \yl{choices}.

\subsection{Datasets and Metrics}

To validate the effectiveness of our method, we performed comprehensive experiments using widely used NeRF-Synthetic~\cite{pumarola2021d} \yl{dataset}, synthetic data from SketchFab~\cite{Sketchfab} (butterfly, Dress, Jeans, Sofa, Digital Human), as well as the real-world scenes that we captured by ourselves.

NeRF-Synthetic \yl{dataset contains} eight static scenes with $360^{\circ}$ random viewpoint settings with the known camera poses.
We use Blender to generate training data for other synthetic data with the identical configuration as NeRF-Synthetic.
The self-captured data is captured by \yl{a} mobile phone, \yl{and} its camera poses are calibrated by \yl{COLMAP}~\cite{schoenberger2016sfm,schoenberger2016mvs}.
All datasets involve a diverse range of visual appearances and geometric properties of different objects that can be deformed.
In order to evaluate the effectiveness of our \yl{approach}, 
we use the three metrics to measure the quality of novel view synthesis, including Peak Signal-to-Noise Ratio (PSNR), Structural Similarity (SSIM~\cite{wang2004image}), and Learned Perceptual Image Patch Similarity (LPIPS~\cite{Zhang2018TheUE}).

\noindent\textbf{Implementation Details}
The pipeline comprises two main components: Gaussian distribution learning using our proposed representation given a set of calibrated images; and performing real-time deforming for 3D Gaussian via an intuitive and friendly `drag' way.
The explicit mesh can be easily extracted by NeuS2~\cite{wang2023neus2} or created by artists (\ie some synthetic data). 
In addition, the utilization of mesh sequences obtained from dynamics or animated meshes can enhance the data-driven deformation of 3D GS, a capability that was not possible with the previous methods.
The training period required for the Gaussian distribution is approximately 10 minutes, using a resolution of 1024$\times$1024. Following this, the 3DGS is deformable in a real-time manner.
Our code is built upon the 3D GS, incorporating effective rendering and training capabilities.
All of the experiments are performed on a computer equipped with an i9-12900K CPU and an RTX 4090 graphics card.
It is important to mention that our pipeline solely focuses on optimizing the Gaussian parameters without any involvement of network parameters.

\begin{table}[htbp]
    \setlength{\tabcolsep}{6pt}
    \setlength{\aboverulesep}{0pt}
    \setlength{\belowrulesep}{0pt}
    \centering
    \caption{
    \textbf{Comparisons of Novel View Synthesis on NeRF-Synthetic Datasets.} 
    We performed the quantitative comparison with three alternative methods on the novel view synthesis task. Three baselines (NeRF-Editing, 3D GS, SuGaR) are evaluated by the three metrics (PSNR, SSIM, LPIPS). The reported results in the table illustrate that our method achieves the best on PSNR and SSIM metrics, and reaches comparable performance on the LPIPS. It demonstrates that our methods can synthesize high-fidelity renderings and support arbitrary edits on 3D GS.
    }
      \begin{tabular}{lcccc}
      \toprule[1.2pt]
      Methods  & NeRF-Editing & 3D-GS & SuGaR & Ours \\
      \midrule
      PSNR$\uparrow$     &   30.58      &   \cellcolor{orange!40}33.31 &   \cellcolor{yellow!40}32.39 & \cellcolor{red!40}33.43 \\
      SSIM$\uparrow$     &   0.960   &   \cellcolor{orange!40}0.967 &   \cellcolor{yellow!40}0.958 & \cellcolor{red!40}0.968 \\
      LPIPS$\downarrow$  &   0.058    &   \cellcolor{red!40}0.023 &   \cellcolor{yellow!40}0.037 & \cellcolor{orange!40}0.034 \\
      \bottomrule[1.2pt]
      \end{tabular}%
    \label{tab:comparison_nvs}%
  \end{table}%

\subsection{Comparisons \& Evaluations}

In this subsection, we evaluate the performance on the above datasets both qualitative and quantitative, including two tasks: novel view synthesis and \yl{deformation}.
We consider three state-of-the-art (SoTA) approaches for these tasks: NeRF-Editing, SuGaR, and 3D-GS.
3D-GS is the baseline for the novel view synthesis, and we selected two classical deformation methods (NeRF-Editing and SuGaR) for NeRF and 3D GS representation.

The quantitative results presented in Table~\ref{tab:comparison_nvs} clearly show that our techniques outperform the three baselines in terms of PSNR and SSIM, while reaching comparable performance in the LPIPS metric.
The results reveal that our approach has the ability to generate realistic results from novel views.

In addition, in order to validate the effectiveness of our methods for deforming, we evaluated the visual results of novel views qualitatively since the deforming results cannot obtain the ground truth.
We perform the experiments on a total of four cases, comprising both synthetic data and real-captured data.
Three alternative methods are evaluated, including NeRF-Editing, SuGaR, and a baseline based on 3D GS.
The baseline is an extension of the 3D GS without incorporating our essential designs (Face Split, Normal Guidance, and regularization $L_r$).
And the Gaussian distribution is constrained to align with the reconstructed surface since there is an explicit surface as guidance.
Next, we deformed the explicit meshes and used the deformed mesh to control the Gaussians using predefined barycentric coordinates.

Fig.~\ref{fig:comparisons} presents four examples that demonstrate our deforming performance through the modification of the explicit mesh.
NeRF-Editing deforms NeRF by utilizing a derived tetrahedron mesh, which is time-consuming and cannot handle large deformations. This often results in blurry rendering outcomes, particularly for high-frequency details.
The baseline is only to attach the Gaussian to the surface without incorporating any special designs.
Consequently, this approach leads to the presence of certain irregularly shaped Gaussian elements, resulting in artifacts when significant deformation and high-frequency features appear.
Although SuGaR successfully reconstructs the mesh from the 3D GS and realizes the deformation by adjusting the parameters of Gaussians, it still fails when capturing fine features at high frequencies during significant deformations (\eg first row in Fig.~\ref{fig:comparisons}).
In contrast to existing methods, our approach effectively models a better Gaussian distribution by using explicit mesh guidance, resulting in enhanced rendering quality and real-time deforming performance at a frame rate of 65 FPS (more deforming results in our videos).

\subsection{More Deformation Results}
Our technique enables the manipulation of 3D GS and an interactive tool that allows for real-time deformation by dragging the sparse control points.
The learned Gaussian distribution is effectively guided by our mesh-based GS, and allows for excellent generalization, even in the face of challenging deformations.
In order to verify the generalizability of our proposed deforming pipeline, we presented more deforming examples involving various deformations in Fig.~\ref{fig:moreresults}.
There are more synthetic examples and real-captured examples.
To obtain more real-time deforming results, please refer to our supplementary video.

\begin{table*}[htbp]
    \setlength{\tabcolsep}{6pt}
    \setlength{\aboverulesep}{0pt}
    \setlength{\belowrulesep}{0pt}
    \centering
    \caption{\textbf{Ablation studies of our key designs on NeRF Synthetic dataset.}
    We evaluate our key designs quantitatively on the NeRF-Synthetic dataset. Three commonly used metrics (PSNR, SSIM, LPIPS) are reported in the table. The results clearly demonstrate that our full method reaches the best performance, and they also indicate that our key designs significantly contribute to our overall pipeline. Note that, \textit{F.S.} means Face split,  and \textit{N.G.} means Normal Guidance.
    }
    \begin{adjustbox}{width=0.85\linewidth}
    \begin{tabular}{lccc|ccc|ccc}
        \toprule[1.2pt]
        \multicolumn{1}{c}{\multirow{2}{*}{Methods}} & \multicolumn{3}{c}{PSNR$\uparrow$}     & \multicolumn{3}{c}{SSIM$\uparrow$}     & \multicolumn{3}{c}{LPIPS$\downarrow$} \\
        \cmidrule{2-10}          & w/o \textit{F.S.} & w/o \textit{N.G.} & Ours  & w/o \textit{F.S.} & w/o \textit{N.G.} & Ours  & w/o \textit{F.S.} & w/o \textit{N.G.}  & Ours \\
        \midrule
        Lego    &  35.61    & 33.48     & \textbf{36.20} & 0.976     & 0.977    & \textbf{0.984} & 0.022     & 0.022    & \textbf{0.015} \\
        Ficus   & 33.33     & 32.09     & \textbf{34.42} & 0.986     & 0.984    & \textbf{0.987} & 0.014     & 0.016    & \textbf{0.014} \\
        Drums   & 27.27     & 27.32     & \textbf{27.85} & 0.962     & 0.959    & \textbf{0.963} & 0.046     & 0.048    & \textbf{0.044} \\
        Chair   & 34.76     & 33.53     & \textbf{35.76} & 0.981     & 0.975    & \textbf{0.986} & 0.017     & 0.020    & \textbf{0.014} \\
        Hotdog  & 38.41     & 36.42     & \textbf{38.45} & 0.963     & 0.962    & \textbf{0.988} & 0.018     & 0.021    & \textbf{0.017} \\
        Mic      & 33.59     & 33.02    & \textbf{34.76} & 0.990     & 0.988    & \textbf{0.991} & 0.008     & 0.009    & \textbf{0.008} \\
        Material & 29.37     & 28.85    & \textbf{29.56} & 0.943     & 0.941    & \textbf{0.956} & 0.044     & 0.046    & \textbf{0.040} \\
        Ship    & 29.47     & 29.36     & \textbf{30.47} & 0.873     & 0.873   & \textbf{0.889} & 0.132     & 0.137    & \textbf{0.124} \\
        \midrule
        Average & 32.73     & 31.76     & \textbf{33.43} & 0.959     & 0.957     & \textbf{0.968} &  0.038    & 0.040    & \textbf{0.034} \\
        \bottomrule[1.2pt]
        \end{tabular}%
    \end{adjustbox}
    \vspace{-2mm}
    \label{tab:abla_nvs}%
\end{table*}%

\subsection{Ablations}
In this subsection, we evaluate various crucial designs in our representation and deforming pipeline by conducting the ablations. These include the Face Split operation, the utilization of Normal Guidance for Gaussian distribution learning, the regularization $L_r$ for 3D GS deforming, and the influence on 3D Gaussian deformation with different resolution of explicit mesh.

\noindent\textbf{Face Split Operation (\textit{F.S.}).}
In order to facilitate the 3DGS deforming, we introduce the explicit mesh as the constraint to bind the Gaussians onto the surface as much as possible.
Consequently, we design a new Gaussian division strategy that aligns with the subdivision of mesh \yl{triangles}.
During each iteration of training, the face will be divided into four smaller faces if it meets the specified criterion (\ie a threshold $2\times10^{-4}$).
Additionally, \yl{each face} will have a Gaussian connected to it.
To validate the effectiveness of face split operation, we remove the operation and instead employ a straightforward division strategy: the Gaussian will be divided into four Gaussians and remain connected to the current face for the ablated version.
We perform the quantitative and qualitative evaluations.
Table~\ref{tab:abla_nvs} reports the scores on the novel view synthesis under the three commonly used metrics (\ie PSNR, SSIM, LPIPS).
Figure~\ref{fig:ablation_sub_loss} incorporates the quantitative results of the 3DGS deformation, which together demonstrate that the face split operation can improve the rendering results from novel views and prevent blurry rendering results after deformation for some high-frequency cases (\eg second and third columns in Figure~\ref{fig:ablation_sub_loss}.).

\noindent\textbf{Normal Guidance (\textit{N.G.}).}
To accurately capture high-frequency details, we propose normal guidance, which ensures that the Gaussian function is positioned in close proximity to the surface rather than just on the surface.
The Gaussian can be translated along the perpendicular direction of the attached face with the assistance of normal guidance, in order to get high-quality rendering results for novel view synthesis.
Table~\ref{tab:abla_nvs} and Figure~\ref{fig:ablation_normal} together illustrate that normal guidance can successfully improve the rendering quality of novel views both quantitatively and qualitatively (\eg the highlighted detailed structures of the Ship and Lego examples).
Note that the normal guidance improves the rendering quality significantly; if we remove it, it also affects the deformation results, so we only validate it on the novel view synthesis task.

\noindent\textbf{Regularization $L_r$.}
Following the training of the Gaussian distribution, an increased number of long-narrow shaped Gaussian shapes are utilized to accurately represent high-frequency appearances. 
However, these Gaussian shapes are not well-suited for deformation, particularly when dealing with large-scale deformations.
To tackle this issue, we introduce the regularization $L_r$ on the size of the Gaussian.
In a word, the loss term will enforce the size of the Gaussian \yl{to be} smaller than the associated faces as much as possible, which achieves better rendering results after the large scale deforming of 3DGS.
In Figure~\ref{fig:ablation_sub_loss}, it is obvious that the ablated version (w/o $L_r$) leads to more artifacts when large-scale deformations appear.
Applying additional constraints to the Gaussians \yl{may have a negative impact on %
rendering results}, \yl{so an appropriate balancing weight should be set to effectively enhance the deformation quality for large-scale deformations while ensuring the novel view synthesis quality.}

\begin{figure}
    \centering
        \includegraphics[width=0.24\linewidth]{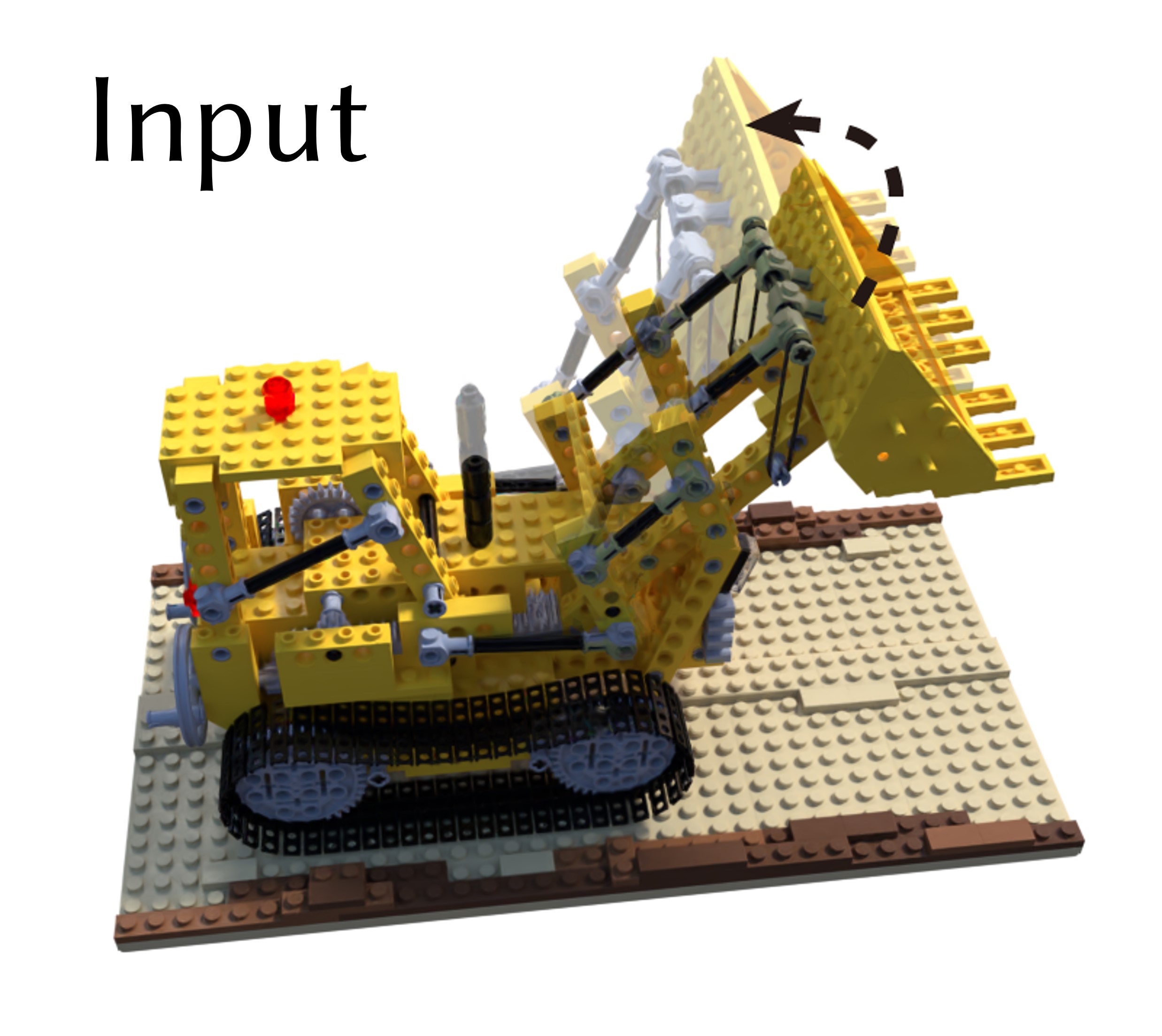}
        \includegraphics[width=0.24\linewidth]{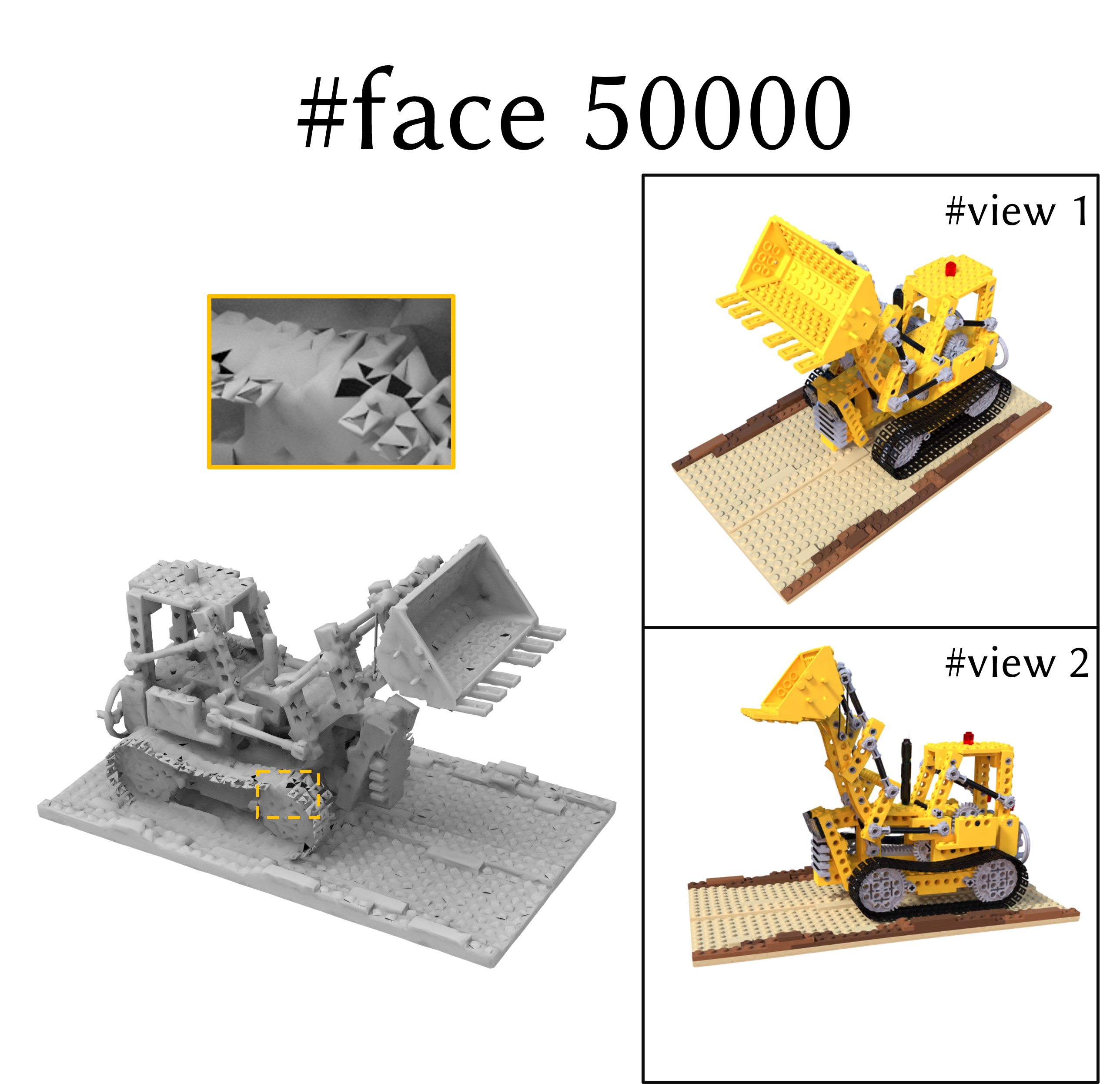}
        \includegraphics[width=0.24\linewidth]{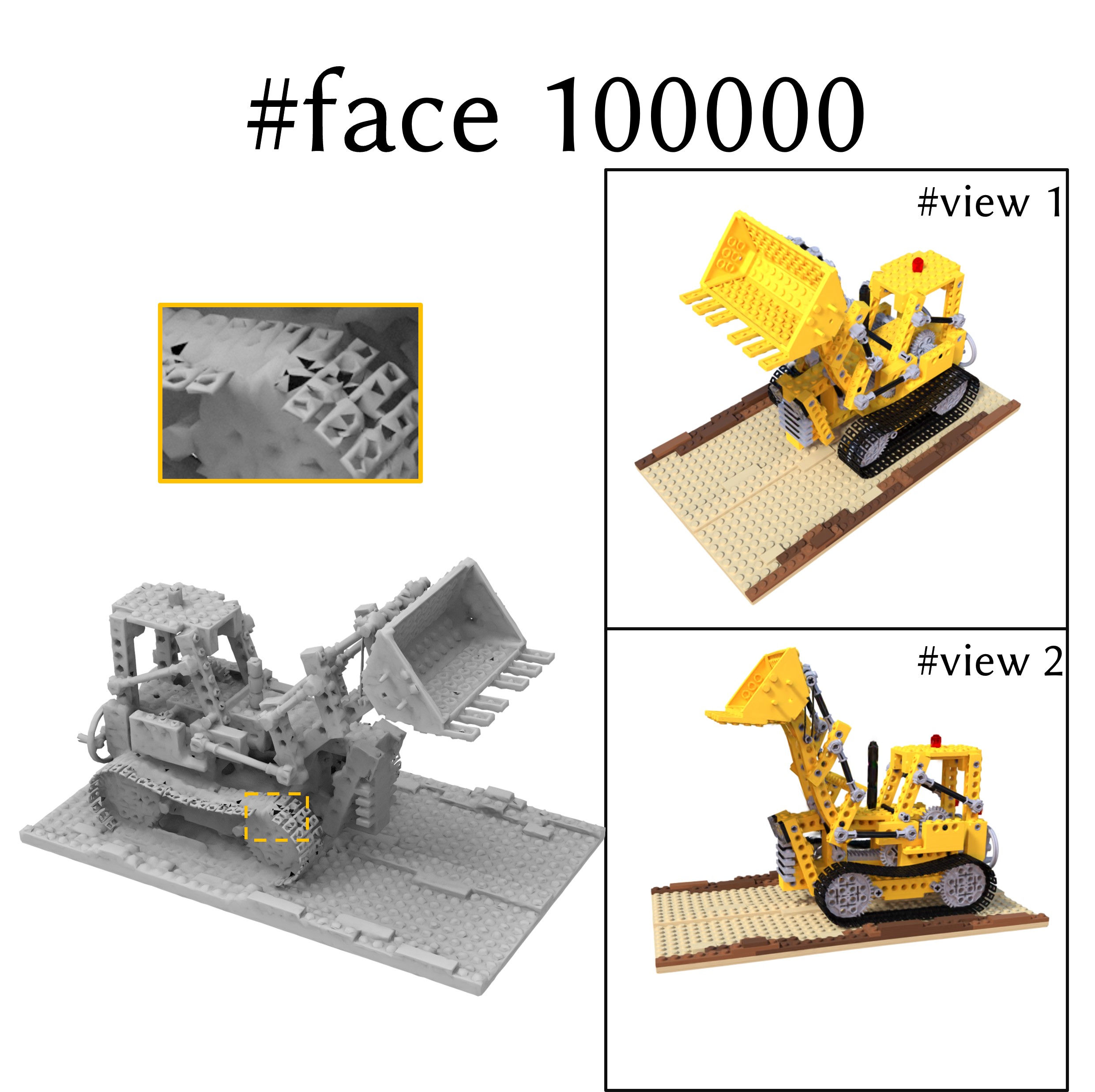}
        \includegraphics[width=0.24\linewidth]{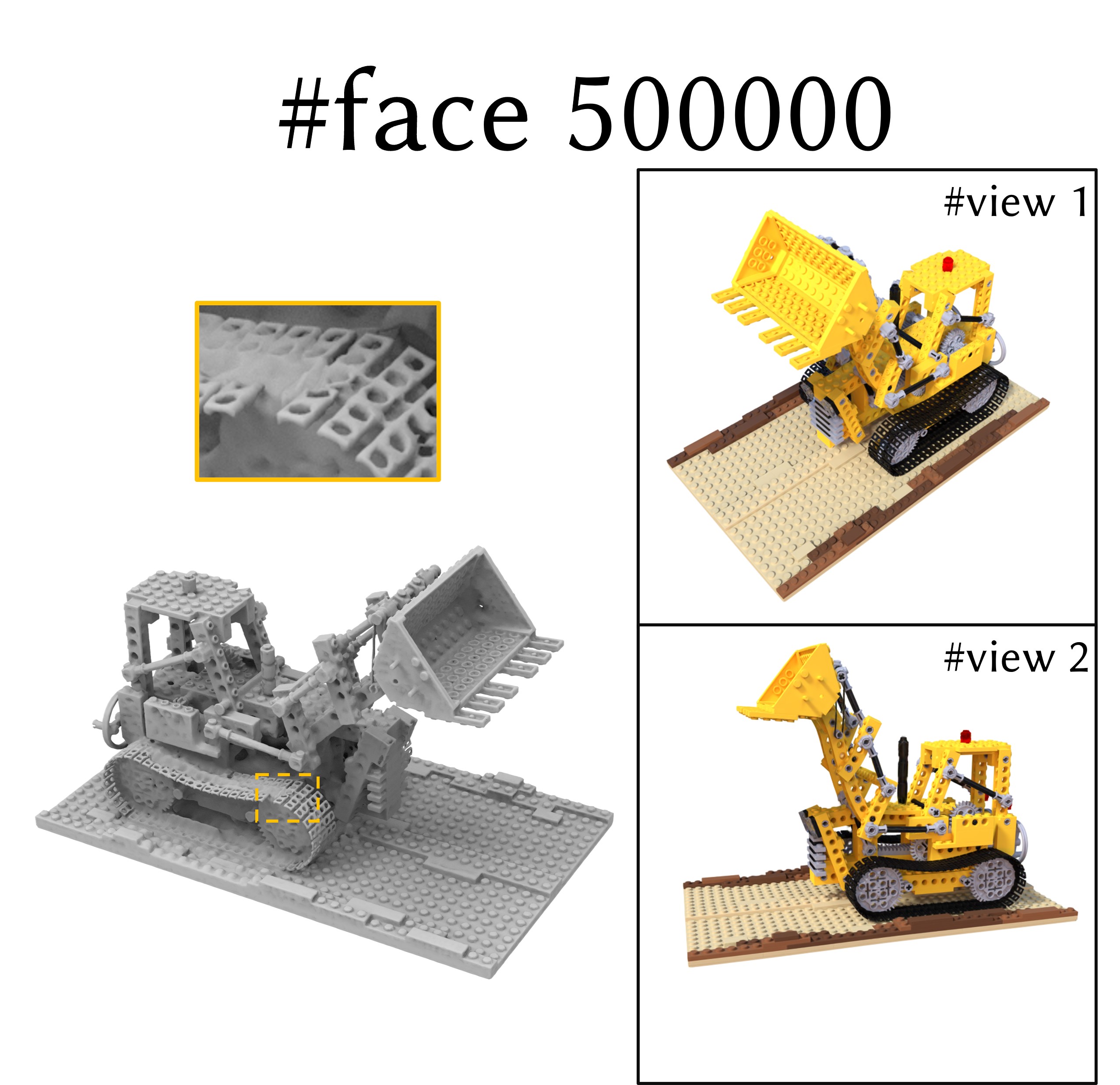}
    \caption{\textbf{Ablation on the explicit mesh with different \yl{resolutions}.}
By employing the explicit mesh as a guide for Gaussian learning, we evaluate the impact of varying mesh resolutions on the deformation results obtained from novel view points. We evaluate the performance of three different mesh resolutions: 50000, 100000, and 500000 \yl{vertices}. The results indicate that our method is not sensitive to the mesh resolution when generating novel views of deformed 3D Gaussians.
    }
    \label{fig:mesh}
\end{figure}

\noindent\textbf{Explicit mesh with different resolutions.}
To verify the influence of mesh resolution on deformed results from novel views, we have simplified the reconstructed mesh using several resolutions, ranging from 500,000 to 50,000 vertices, as explicit guidance to drive the 3D Gaussian deformation. 
Figure~\ref{fig:mesh} demonstrates that our 3D Gaussian deformation pipeline is not sensitive to the mesh resolution, which successfully achieves similar performance on Gaussian deformation from novel views.

\section{Conclusions \& Future Work}\label{sec:con}
\subsection{Conclusions}
In this paper, we \yl{have proposed} a novel large scale deformation method for 3D Gaussian Splatting, based on \yl{a} mesh-based representation.
\yl{The} proposed 3DGS deformation method enables the manipulating of the 3DGS in an intuitive interactive manner.
To well facilitate the 3DGS deformation, we incorporate \yl{an explicit mesh that can be} easily extracted by existing methods, which \yl{is} bound with Gaussian ellipsoids together and \yl{enables the effective large-scale deformation} of 3DGS.
In addition, we employ a Gaussian division model that operates on the explicit mesh through face split and normal guidance, which can improve the visual quality and prevents artifacts that may occur during large-scale deformations.
Based on the mesh-based GS representation, we introduce a large-scale mesh deformation method to enable deformable Gaussian Splatting by altering the parameters of the 3D Gaussians according to the user's intuitive manipulations.

\subsection{Limitations \& Future Work}
Despite this approach successfully achieving real-time large-scale deformation with Gaussian Splatting representation, it still faces the following obstacles:
1) The visual appearance and shadow are still baked on the Gaussians and cannot be further edited. 
2) This method relies on the extracted mesh as the proxy, it would fail if the mesh could not be extracted such as for complex transparent objects. 
In future work, it is worth developing new methods that can not only deform the geometry of Gaussians, but also support editing of the appearance of Gaussians. 
In addition, our representation could also be applied to other applications such as digital human avatars, and further development for novel applications will be explored as future work.

\ifCLASSOPTIONcaptionsoff
  \newpage
\fi

{
\bibliographystyle{IEEEtran}
\bibliography{bibliography.bib}

\begin{thebibliography}{10}
\providecommand{\url}[1]{#1}
\csname url@samestyle\endcsname
\providecommand{\newblock}{\relax}
\providecommand{\bibinfo}[2]{#2}
\providecommand{\BIBentrySTDinterwordspacing}{\spaceskip=0pt\relax}
\providecommand{\BIBentryALTinterwordstretchfactor}{4}
\providecommand{\BIBentryALTinterwordspacing}{\spaceskip=\fontdimen2\font plus
\BIBentryALTinterwordstretchfactor\fontdimen3\font minus
  \fontdimen4\font\relax}
\providecommand{\BIBforeignlanguage}[2]{{%
\expandafter\ifx\csname l@#1\endcsname\relax
\typeout{** WARNING: IEEEtran.bst: No hyphenation pattern has been}%
\typeout{** loaded for the language `#1'. Using the pattern for}%
\typeout{** the default language instead.}%
\else
\language=\csname l@#1\endcsname
\fi
#2}}
\providecommand{\BIBdecl}{\relax}
\BIBdecl

\bibitem{kerbl20233d}
B.~Kerbl, G.~Kopanas, T.~Leimk{\"u}hler, and G.~Drettakis, ``3d gaussian
  splatting for real-time radiance field rendering,'' \emph{ACM Transactions on
  Graphics (ToG)}, vol.~42, no.~4, pp. 1--14, 2023.

\bibitem{qi2017pointnet}
C.~R. Qi, H.~Su, K.~Mo, and L.~J. Guibas, ``{PointNet}: Deep learning on point
  sets for {3D} classification and segmentation,'' in \emph{Proceedings of the
  IEEE conference on computer vision and pattern recognition}, 2017, pp.
  652--660.

\bibitem{qi2017pointnet++}
C.~R. Qi, L.~Yi, H.~Su, and L.~J. Guibas, ``{PointNet++}: Deep hierarchical
  feature learning on point sets in a metric space,'' in \emph{Advances in
  neural information processing systems}, 2017, pp. 5099--5108.

\bibitem{fan2017point}
H.~Fan, H.~Su, and L.~J. Guibas, ``A point set generation network for {3D}
  object reconstruction from a single image,'' in \emph{Proceedings of the IEEE
  conference on computer vision and pattern recognition}, 2017, pp. 605--613.

\bibitem{achlioptas2018learning}
P.~Achlioptas, O.~Diamanti, I.~Mitliagkas, and L.~Guibas, ``Learning
  representations and generative models for {3D} point clouds,'' in
  \emph{ICML}, 2018, pp. 40--49.

\bibitem{maturana2015voxnet}
D.~Maturana and S.~Scherer, ``{VoxNet}: A {3D} convolutional neural network for
  real-time object recognition,'' in \emph{2015 IEEE/RSJ International
  Conference on Intelligent Robots and Systems (IROS)}.\hskip 1em plus 0.5em
  minus 0.4em\relax IEEE, 2015, pp. 922--928.

\bibitem{choy20163d}
C.~B. Choy, D.~Xu, J.~Gwak, K.~Chen, and S.~Savarese, ``{3D-R2N2}: A unified
  approach for single and multi-view {3D} object reconstruction,'' in
  \emph{ECCV}.\hskip 1em plus 0.5em minus 0.4em\relax Springer, 2016, pp.
  628--644.

\bibitem{3dgan}
J.~Wu, C.~Zhang, T.~Xue, W.~T. Freeman, and J.~B. Tenenbaum, ``Learning a
  probabilistic latent space of object shapes via 3d generative-adversarial
  modeling,'' in \emph{Advances in Neural Information Processing Systems},
  2016, pp. 82--90.

\bibitem{wang2018pixel2mesh}
N.~Wang, Y.~Zhang, Z.~Li, Y.~Fu, W.~Liu, and Y.-G. Jiang, ``Pixel2mesh:
  Generating {3D} mesh models from single {RGB} images,'' in \emph{Proceedings
  of the European Conference on Computer Vision (ECCV)}, 2018, pp. 52--67.

\bibitem{groueix2018atlasnet}
T.~Groueix, M.~Fisher, V.~G. Kim, B.~C. Russell, and M.~Aubry, ``{AtlasNet}: A
  papier-m\^ach\'e approach to learning {3D} surface generation,'' in
  \emph{Proceedings of the IEEE Conference on Computer Vision and Pattern
  Recognition}, 2018.

\bibitem{mildenhall2021nerf}
B.~Mildenhall, P.~P. Srinivasan, M.~Tancik, J.~T. Barron, R.~Ramamoorthi, and
  R.~Ng, ``Nerf: Representing scenes as neural radiance fields for view
  synthesis,'' \emph{Communications of the ACM}, vol.~65, no.~1, pp. 99--106,
  2021.

\bibitem{barron2022mip}
J.~T. Barron, B.~Mildenhall, D.~Verbin, P.~P. Srinivasan, and P.~Hedman,
  ``Mip-nerf 360: Unbounded anti-aliased neural radiance fields,'' in
  \emph{Proceedings of the IEEE/CVF Conference on Computer Vision and Pattern
  Recognition}, 2022, pp. 5470--5479.

\bibitem{mescheder2019occupancy}
L.~Mescheder, M.~Oechsle, M.~Niemeyer, S.~Nowozin, and A.~Geiger, ``Occupancy
  networks: Learning {3D} reconstruction in function space,'' in
  \emph{Proceedings of the IEEE Conference on Computer Vision and Pattern
  Recognition}, 2019, pp. 4460--4470.

\bibitem{chibane2020implicit}
J.~Chibane, T.~Alldieck, and G.~Pons-Moll, ``Implicit functions in feature
  space for 3d shape reconstruction and completion,'' in \emph{Proceedings of
  the IEEE/CVF Conference on Computer Vision and Pattern Recognition}, 2020,
  pp. 6970--6981.

\bibitem{chen2019learning}
Z.~Chen and H.~Zhang, ``Learning implicit fields for generative shape
  modeling,'' in \emph{Proceedings of the IEEE Conference on Computer Vision
  and Pattern Recognition}, 2019, pp. 5939--5948.

\bibitem{ullman1979interpretation}
S.~Ullman, ``The interpretation of structure from motion,'' \emph{Proceedings
  of the Royal Society of London. Series B. Biological Sciences}, vol. 203, no.
  1153, pp. 405--426, 1979.

\bibitem{huang2023sc}
Y.-H. Huang, Y.-T. Sun, Z.~Yang, X.~Lyu, Y.-P. Cao, and X.~Qi, ``Sc-gs:
  Sparse-controlled gaussian splatting for editable dynamic scenes,''
  \emph{arXiv preprint arXiv:2312.14937}, 2023.

\bibitem{guedon2023sugar}
A.~Gu{\'e}don and V.~Lepetit, ``Sugar: Surface-aligned gaussian splatting for
  efficient 3d mesh reconstruction and high-quality mesh rendering,''
  \emph{arXiv preprint arXiv:2311.12775}, 2023.

\bibitem{wang2023neus2}
Y.~Wang, Q.~Han, M.~Habermann, K.~Daniilidis, C.~Theobalt, and L.~Liu, ``Neus2:
  Fast learning of neural implicit surfaces for multi-view reconstruction,'' in
  \emph{Proceedings of the IEEE/CVF International Conference on Computer
  Vision}, 2023, pp. 3295--3306.

\bibitem{gao2019sparse}
L.~Gao, Y.-K. Lai, J.~Yang, L.-X. Zhang, S.~Xia, and L.~Kobbelt, ``Sparse data
  driven mesh deformation,'' \emph{IEEE transactions on visualization and
  computer graphics}, vol.~27, no.~3, pp. 2085--2100, 2019.

\bibitem{wang2018adaptive}
P.-S. Wang, C.-Y. Sun, Y.~Liu, and X.~Tong, ``Adaptive o-cnn: A patch-based
  deep representation of 3d shapes,'' \emph{ACM Transactions on Graphics
  (TOG)}, vol.~37, no.~6, pp. 1--11, 2018.

\bibitem{hanocka2019meshcnn}
R.~Hanocka, A.~Hertz, N.~Fish, R.~Giryes, S.~Fleishman, and D.~Cohen-Or,
  ``Meshcnn: a network with an edge,'' \emph{ACM Transactions on Graphics
  (TOG)}, vol.~38, no.~4, pp. 1--12, 2019.

\bibitem{gaosdmnet2019}
L.~Gao, J.~Yang, T.~Wu, Y.-J. Yuan, H.~Fu, Y.-K. Lai, and H.~Zhang,
  ``{SDM-NET}: Deep generative network for structured deformable mesh,''
  \emph{ACM Transactions on Graphics (Proceedings of ACM SIGGRAPH Asia 2019)},
  vol.~38, no.~6, pp. 243:1--243:15, 2019.

\bibitem{yang2022dsg}
J.~Yang, K.~Mo, Y.-K. Lai, L.~J. Guibas, and L.~Gao, ``Dsg-net: Learning
  disentangled structure and geometry for 3d shape generation,'' \emph{ACM
  Transactions on Graphics (TOG)}, vol.~42, no.~1, pp. 1--17, 2022.

\bibitem{chibane2020neural}
J.~Chibane, A.~Mir, and G.~Pons-Moll, ``Neural unsigned distance fields for
  implicit function learning,'' \emph{arXiv preprint arXiv:2010.13938}, 2020.

\bibitem{guillard2021meshudf}
B.~Guillard, F.~Stella, and P.~Fua, ``Meshudf: Fast and differentiable meshing
  of unsigned distance field networks,'' \emph{arXiv preprint
  arXiv:2111.14549}, 2021.

\bibitem{liu2023neudf}
Y.-T. Liu, L.~Wang, J.~Yang, W.~Chen, X.~Meng, B.~Yang, and L.~Gao, ``Neudf:
  Leaning neural unsigned distance fields with volume rendering,'' in
  \emph{Proceedings of the IEEE/CVF Conference on Computer Vision and Pattern
  Recognition}, 2023, pp. 237--247.

\bibitem{kajiya1984ray}
J.~T. Kajiya and B.~P. Von~Herzen, ``Ray tracing volume densities,'' \emph{ACM
  SIGGRAPH computer graphics}, vol.~18, no.~3, pp. 165--174, 1984.

\bibitem{li2023neuralangelo}
Z.~Li, T.~M\"uller, A.~Evans, R.~H. Taylor, M.~Unberath, M.-Y. Liu, and C.-H.
  Lin, ``Neuralangelo: High-fidelity neural surface reconstruction,'' in
  \emph{CVPR}, 2023.

\bibitem{wang2021neus}
P.~Wang, L.~Liu, Y.~Liu, C.~Theobalt, T.~Komura, and W.~Wang, ``Neus: Learning
  neural implicit surfaces by volume rendering for multi-view reconstruction,''
  \emph{arXiv preprint arXiv:2106.10689}, 2021.

\bibitem{poole2022dreamfusion}
B.~Poole, A.~Jain, J.~T. Barron, and B.~Mildenhall, ``Dreamfusion: Text-to-3d
  using 2d diffusion,'' \emph{arXiv preprint arXiv:2209.14988}, 2022.

\bibitem{liu2023zero}
R.~Liu, R.~Wu, B.~Van~Hoorick, P.~Tokmakov, S.~Zakharov, and C.~Vondrick,
  ``Zero-1-to-3: Zero-shot one image to 3d object,'' in \emph{Proceedings of
  the IEEE/CVF International Conference on Computer Vision}, 2023, pp.
  9298--9309.

\bibitem{haque2023instruct}
A.~Haque, M.~Tancik, A.~A. Efros, A.~Holynski, and A.~Kanazawa,
  ``Instruct-nerf2nerf: Editing 3d scenes with instructions,'' \emph{arXiv
  preprint arXiv:2303.12789}, 2023.

\bibitem{wang2023seal}
X.~Wang, J.~Zhu, Q.~Ye, Y.~Huo, Y.~Ran, Z.~Zhong, and J.~Chen, ``Seal-3d:
  Interactive pixel-level editing for neural radiance fields,'' in
  \emph{Proceedings of the IEEE/CVF International Conference on Computer
  Vision}, 2023, pp. 17\,683--17\,693.

\bibitem{liu2021editing}
S.~Liu, X.~Zhang, Z.~Zhang, R.~Zhang, J.-Y. Zhu, and B.~Russell, ``Editing
  conditional radiance fields,'' in \emph{Proceedings of the IEEE/CVF
  international conference on computer vision}, 2021, pp. 5773--5783.

\bibitem{muller2022instant}
T.~M{\"u}ller, A.~Evans, C.~Schied, and A.~Keller, ``Instant neural graphics
  primitives with a multiresolution hash encoding,'' \emph{ACM Transactions on
  Graphics (ToG)}, vol.~41, no.~4, pp. 1--15, 2022.

\bibitem{yu2021plenoctrees}
A.~Yu, R.~Li, M.~Tancik, H.~Li, R.~Ng, and A.~Kanazawa, ``Plenoctrees for
  real-time rendering of neural radiance fields,'' in \emph{Proceedings of the
  IEEE/CVF International Conference on Computer Vision}, 2021, pp. 5752--5761.

\bibitem{fridovich2022plenoxels}
S.~Fridovich-Keil, A.~Yu, M.~Tancik, Q.~Chen, B.~Recht, and A.~Kanazawa,
  ``Plenoxels: Radiance fields without neural networks,'' in \emph{Proceedings
  of the IEEE/CVF Conference on Computer Vision and Pattern Recognition}, 2022,
  pp. 5501--5510.

\bibitem{chen2024survey}
G.~Chen and W.~Wang, ``A survey on 3d gaussian splatting,'' \emph{arXiv
  preprint arXiv:2401.03890}, 2024.

\bibitem{luiten2023dynamic}
J.~Luiten, G.~Kopanas, B.~Leibe, and D.~Ramanan, ``Dynamic 3d gaussians:
  Tracking by persistent dynamic view synthesis,'' in \emph{3DV}, 2024.

\bibitem{yang2023deformable3dgs}
Z.~Yang, X.~Gao, W.~Zhou, S.~Jiao, Y.~Zhang, and X.~Jin, ``Deformable 3d
  gaussians for high-fidelity monocular dynamic scene reconstruction,''
  \emph{arXiv preprint arXiv:2309.13101}, 2023.

\bibitem{li2023animatable}
Z.~Li, Z.~Zheng, L.~Wang, and Y.~Liu, ``Animatable gaussians: Learning
  pose-dependent gaussian maps for high-fidelity human avatar modeling,''
  \emph{arXiv preprint arXiv:2311.16096}, 2023.

\bibitem{kocabas2023hugs}
M.~Kocabas, J.-H.~R. Chang, J.~Gabriel, O.~Tuzel, and A.~Ranjan, ``Hugs: Human
  gaussian splats,'' \emph{arXiv preprint arXiv:2311.17910}, 2023.

\bibitem{chen2023text}
Z.~Chen, F.~Wang, and H.~Liu, ``Text-to-3d using gaussian splatting,''
  \emph{arXiv preprint arXiv:2309.16585}, 2023.

\bibitem{tang2023dreamgaussian}
J.~Tang, J.~Ren, H.~Zhou, Z.~Liu, and G.~Zeng, ``Dreamgaussian: Generative
  gaussian splatting for efficient 3d content creation,'' \emph{arXiv preprint
  arXiv:2309.16653}, 2023.

\bibitem{yi2023gaussiandreamer}
T.~Yi, J.~Fang, G.~Wu, L.~Xie, X.~Zhang, W.~Liu, Q.~Tian, and X.~Wang,
  ``Gaussiandreamer: Fast generation from text to 3d gaussian splatting with
  point cloud priors,'' \emph{arXiv preprint arXiv:2310.08529}, 2023.

\bibitem{Lipman2005LaplacianFF}
Y.~Lipman, O.~Sorkine-Hornung, M.~Alexa, D.~Cohen-Or, D.~Levin, C.~R{\"o}ssl,
  and H.-P. Seidel, ``Laplacian framework for interactive mesh editing,''
  \emph{Int. J. Shape Model.}, vol.~11, pp. 43--62, 2005.

\bibitem{SorkineHornung2005LaplacianMP}
O.~Sorkine-Hornung, ``Laplacian mesh processing,'' in \emph{Eurographics},
  2005.

\bibitem{sorkine2007rigid}
O.~Sorkine and M.~Alexa, ``As-rigid-as-possible surface modeling,'' in
  \emph{Symposium on Geometry processing}, vol.~4.\hskip 1em plus 0.5em minus
  0.4em\relax Citeseer, 2007, pp. 109--116.

\bibitem{yu2004mesh}
Y.~Yu, K.~Zhou, D.~Xu, X.~Shi, H.~Bao, B.~Guo, and H.-Y. Shum, ``Mesh editing
  with {Poisson-based} gradient field manipulation,'' in \emph{ACM SIGGRAPH
  2004 Papers}, 2004, pp. 644--651.

\bibitem{wang2020NeuralCage}
W.~Yifan, N.~Aigerman, V.~G. Kim, S.~Chaudhuri, and O.~Sorkine-Hornung,
  ``Neural cages for detail-preserving {3D} deformations,'' in
  \emph{Proceedings of the IEEE/CVF Conference on Computer Vision and Pattern
  Recognition}, 2020, pp. 75--83.

\bibitem{zhang2020proxy}
Y.~Zhang, J.~Zheng, and Y.~Cai, ``Proxy-driven free-form deformation by
  topology-adjustable control lattice,'' \emph{Computers \& Graphics}, vol.~89,
  pp. 167--177, 2020.

\bibitem{sumner2005mesh}
R.~W. Sumner, M.~Zwicker, C.~Gotsman, and J.~Popovi{\'c}, ``Mesh-based inverse
  kinematics,'' \emph{ACM transactions on graphics (TOG)}, vol.~24, no.~3, pp.
  488--495, 2005.

\bibitem{wang2022clip}
C.~Wang, M.~Chai, M.~He, D.~Chen, and J.~Liao, ``Clip-nerf: Text-and-image
  driven manipulation of neural radiance fields,'' in \emph{Proceedings of the
  IEEE/CVF Conference on Computer Vision and Pattern Recognition}, 2022, pp.
  3835--3844.

\bibitem{wang2023nerf}
C.~Wang, R.~Jiang, M.~Chai, M.~He, D.~Chen, and J.~Liao, ``Nerf-art:
  Text-driven neural radiance fields stylization,'' \emph{IEEE Transactions on
  Visualization and Computer Graphics}, 2023.

\bibitem{gao2023textdeformer}
W.~Gao, N.~Aigerman, T.~Groueix, V.~G. Kim, and R.~Hanocka, ``Textdeformer:
  Geometry manipulation using text guidance,'' \emph{arXiv preprint
  arXiv:2304.13348}, 2023.

\bibitem{bao2023sine}
C.~Bao, Y.~Zhang, and B.~e.~a. Yang, ``Sine: Semantic-driven image-based nerf
  editing with prior-guided editing field,'' in \emph{CVPR 2023}, 2023, pp.
  20\,919--20\,929.

\bibitem{peng2021neural}
S.~Peng, Y.~Zhang, Y.~Xu, and et~al., ``Neural body: Implicit neural
  representations with structured latent codes for novel view synthesis of
  dynamic humans,'' in \emph{CVPR 2021}, 2021, pp. 9054--9063.

\bibitem{jiang2023avatarcraft}
R.~Jiang, C.~Wang, J.~Zhang, M.~Chai, M.~He, D.~Chen, and J.~Liao,
  ``Avatarcraft: Transforming text into neural human avatars with parameterized
  shape and pose control,'' \emph{arXiv preprint arXiv:2303.17606}, 2023.

\bibitem{Yuan2022NeRFEditingGE}
Y.-J. Yuan, Y.-T. Sun, Y.-K. Lai, Y.~Ma, R.~Jia, and L.~Gao, ``Nerf-editing:
  Geometry editing of neural radiance fields,'' 2022, pp. 18\,332--18\,343.

\bibitem{yang2022neumesh}
B.~Yang, C.~Bao, J.~Zeng, H.~Bao, Y.~Zhang, Z.~Cui, and G.~Zhang, ``Neumesh:
  Learning disentangled neural mesh-based implicit field for geometry and
  texture editing,'' in \emph{European Conference on Computer Vision}.\hskip
  1em plus 0.5em minus 0.4em\relax Springer, 2022, pp. 597--614.

\bibitem{xu2022deforming}
T.~Xu and T.~Harada, ``Deforming radiance fields with cages,'' in
  \emph{European Conference on Computer Vision}.\hskip 1em plus 0.5em minus
  0.4em\relax Springer, 2022, pp. 159--175.

\bibitem{jambon2023nerfshop}
C.~Jambon, B.~Kerbl, G.~Kopanas, S.~Diolatzis, G.~Drettakis, and
  T.~Leimk{\"u}hler, ``Nerfshop: Interactive editing of neural radiance
  fields,'' \emph{Proceedings of the ACM on Computer Graphics and Interactive
  Techniques}, vol.~6, no.~1, 2023.

\bibitem{liu2022nerf}
H.-K. Liu, I.~Shen, B.-Y. Chen \emph{et~al.}, ``Nerf-in: Free-form nerf
  inpainting with rgb-d priors,'' \emph{arXiv preprint arXiv:2206.04901}, 2022.

\bibitem{kobayashi2022decomposing}
S.~Kobayashi, E.~Matsumoto, and V.~Sitzmann, ``Decomposing nerf for editing via
  feature field distillation,'' \emph{arXiv preprint arXiv:2205.15585}, 2022.

\bibitem{zhuang2023dreameditor}
J.~Zhuang, C.~Wang, L.~Liu, L.~Lin, and G.~Li, ``Dreameditor: Text-driven 3d
  scene editing with neural fields,'' \emph{arXiv preprint arXiv:2306.13455},
  2023.

\bibitem{pumarola2021d}
A.~Pumarola, E.~Corona, G.~Pons-Moll, and F.~Moreno-Noguer, ``D-nerf: Neural
  radiance fields for dynamic scenes,'' in \emph{Proceedings of the IEEE/CVF
  Conference on Computer Vision and Pattern Recognition}, 2021, pp.
  10\,318--10\,327.

\bibitem{jiang20234d}
D.~Jiang, Z.~Ke, X.~Zhou, and X.~Shi, ``4d-editor: Interactive object-level
  editing in dynamic neural radiance fields via semantic distillation,''
  \emph{arXiv e-prints}, pp. arXiv--2310, 2023.

\bibitem{xie2023physgaussian}
T.~Xie, Z.~Zong, Y.~Qiu, X.~Li, Y.~Feng, Y.~Yang, and C.~Jiang, ``Physgaussian:
  Physics-integrated 3d gaussians for generative dynamics,'' \emph{arXiv
  preprint arXiv:2311.12198}, 2023.

\bibitem{bonet1997nonlinear}
J.~Bonet and R.~D. Wood, \emph{Nonlinear continuum mechanics for finite element
  analysis}.\hskip 1em plus 0.5em minus 0.4em\relax Cambridge university press,
  1997.

\bibitem{zwicker2001surfacesplatting}
M.~Zwicker, H.~Pfister, J.~Van~Baar, and M.~Gross, ``Surface splatting,'' in
  \emph{Proceedings of the 28th annual conference on Computer graphics and
  interactive techniques}, 2001, pp. 371--378.

\bibitem{Sketchfab}
C.~Pinson, ``Sketchfab - the best 3d viewer on the web,''
  \url{https://sketchfab.com/}, 2011.

\bibitem{deepcloth_su2022}
Z.~Su, T.~Yu, Y.~Wang, and Y.~Liu, ``Deepcloth: Neural garment representation
  for shape and style editing,'' \emph{IEEE Transactions on Pattern Analysis
  and Machine Intelligence}, vol.~45, no.~2, pp. 1581--1593, 2023.

\bibitem{schoenberger2016sfm}
J.~L. Sch\"{o}nberger and J.-M. Frahm, ``Structure-from-motion revisited,'' in
  \emph{Conference on Computer Vision and Pattern Recognition (CVPR)}, 2016.

\bibitem{schoenberger2016mvs}
J.~L. Sch\"{o}nberger, E.~Zheng, M.~Pollefeys, and J.-M. Frahm, ``Pixelwise
  view selection for unstructured multi-view stereo,'' in \emph{European
  Conference on Computer Vision (ECCV)}, 2016.

\bibitem{wang2004image}
Z.~Wang, A.~C. Bovik, H.~R. Sheikh, and E.~P. Simoncelli, ``Image quality
  assessment: from error visibility to structural similarity,'' \emph{IEEE
  transactions on image processing}, vol.~13, no.~4, pp. 600--612, 2004.

\bibitem{Zhang2018TheUE}
R.~Zhang, P.~Isola, A.~A. Efros, E.~Shechtman, and O.~Wang, ``The unreasonable
  effectiveness of deep features as a perceptual metric,'' 2018, pp. 586--595.

\end{thebibliography}
}

\end{document}